\documentclass[notitlepage,a4paper,aps,prd,onecolumn,superscriptaddress,nofootinbib,groupedaddress]{revtex4}
\usepackage{enumerate}
\usepackage{amsmath}
\usepackage{amsfonts}
\usepackage{amssymb}
\usepackage[utf8]{inputenc}
\usepackage[T1]{fontenc}
\usepackage{mathtools}
\usepackage{wasysym}
\usepackage{accents,float}
\usepackage{ulem,color,cancel,soul}

\newcounter{subeq}

\newcommand{\mysubeq}

\DeclareMathAlphabet{\mathds}{U}{BOONDOX-ds}{m}{n}
\usepackage[dvipsnames]{xcolor}

\usepackage{hyperref}
\hypersetup{
    colorlinks=true,
    linkcolor=blue,
    filecolor=magenta,      
   citecolor=blue
}
\usepackage{amsthm}

\theoremstyle{definition}

\theoremstyle{plain}

\usepackage{physics}
\usepackage{bbm}

\allowdisplaybreaks

\begin{document}
\title{Trace Anomaly in Metric-Affine gravity}

\author{Sebastian Bahamonde}
\email{sbahamondebeltran@gmail.com, sebastian.bahamonde@ipmu.jp}
\affiliation{Kavli Institute for the Physics and Mathematics of the Universe (WPI), The University of Tokyo Institutes
for Advanced Study (UTIAS), The University of Tokyo, Kashiwa, Chiba 277-8583, Japan.}

\author{Yuichi Miyashita}
\email{miyashita.y.ae@m.titech.ac.jp}
\affiliation{Department of Physics, Tokyo Institute of Technology 1-12-1 Ookayama, Meguro-ku, Tokyo 152-8551, Japan.}

\author{Masahide Yamaguchi}
\email{gucci@ibs.re.kr}
\affiliation{Cosmology, Gravity, and Astroparticle Physics Group,
Center for Theoretical Physics of the Universe,
Institute for Basic Science (IBS), Daejeon, 34126, Korea}
\affiliation{Department of Physics, Tokyo Institute of Technology 1-12-1 Ookayama, Meguro-ku, Tokyo 152-8551, Japan.}

\begin{abstract}
We explore the trace (Weyl) anomaly within a general metric-affine geometry that includes both torsion and nonmetricity. Using the Heat Kernel method and Seeley's algorithm, we compute the Minakshisundaram coefficients for arbitrary spacetimes within this framework, incorporating the effects of the nonmetricity and torsion tensors for the first time. We then determine the corrections to the trace anomaly at one loop for the matter sector in theories invariant under conformal transformation, frame rescaling transformation, and projective transformation. We identify a new anomaly related to hypermomentum, arising from the dilation part mediated by the Weyl component of nonmetricity. As particular cases, we analyze the spin $0$ and spin $1/2$ cases, considering various couplings between matter and the gravitational sector. We demonstrate that invariance under the frame rescaling transformation results in an anomaly in the relationship between the hypermomentum and the stress-energy tensor. In contrast, under the projective transformation, no anomaly is present; specifically, there is no non-zero trace of the hypermomentum tensor in any of our concrete examples.
\end{abstract}

\maketitle
\section{Introduction}\label{sec:intro}

Trace anomaly of an energy momentum tensor \cite{Capper:1974ic} is one of anomalies in quantum field theory.  Under the Weyl symmetry, which also implies the conformal symmetry in the Minkowski background for a diffeomorphism invariant theory \cite{zumino1970lectures}, the trace of the energy momentum tensor vanishes classically. However, once quantum effects are taken into account, its expectation value takes a non-zero value in general, which is called trace anomaly or conformal anomaly \cite{Duff:1993wm,Deser:1996na}. This effect has a lot of interesting applications in particle physics~\cite{Fradkin:1983tg,Riegert:1984kt,Komargodski:2011vj}, cosmology~\cite{Olive:1989nu,Hawking:2000bb,Starobinsky:1980te,Netto:2015cba,Fischetti:1979ue}, and gravity~\cite{Christensen:1977jc,Camargo:2022gcw,Mottola:2022tcn,Komargodski:2011xv}.

The form of the trace anomaly can be classified into two parts. One of them is sourced by spacetime curvature and is often called Weyl anomaly while the other is sourced by local (four dimension) operators and the running of the couplings. In this paper, we concentrate on the former without external gauge fields. The form of trace (Weyl) anomaly depends on the gravity theory and matter content. In the context of the so-called metric formalism of gravity, or equivalently, in the framework of Riemannian geometry, an affine connection is uniquely fixed to be the Levi-Civita one. In such a framework, the form of the Weyl anomaly can be easily calculated and known to consist of only two terms in four dimensions \cite{Duff:1977ay} given by  $\langle T^{\mu}{}_\mu \rangle=a_1 \left(C^2+\frac{2}{3}\Box R\right)+a_2G$ with the Weyl squared term $C^2\equiv C_{\alpha\beta\mu\nu}C^{\alpha\beta\mu\nu}$ and the Gauss-Bonnet term $G=R^2-4R_{\mu\nu}R^{\mu\nu}+R_{\alpha\beta\mu\nu}R^{\alpha\beta\mu\nu}$, and whose coefficients $a_1, a_2$ depend on the matter content. 

Beyond the traditional Riemannian geometry, there are other formalisms of gravity, such as Einstein-Cartan gravity, which employs Riemann-Cartan geometry with torsion, and the more general Metric-Affine Gravity (MAG), which incorporates both torsion and nonmetricity tensors~\cite{Hehl:1994ue,Hehl:1976kj,Obukhov:1987tz,Blagojevic:2013xpa,Obukhov:2022khx}. In both cases, the affine connection is not necessarily the Levi-Civita connection and is only determined after the action is specified, provided these fields are non-dynamical. However, if they possess additional dynamics, they lead to extra field equations obtained from the variation of the action with respect to the affine connection (or the spin connection in the gauge formalism). These theories can also introduce new dynamics and effects in astrophysical and cosmological scenarios~\cite{adamowicz1980plane,Baekler:1981lkh,Gonner:1984rw,Bakler:1988nq,Gladchenko:1994wu,Tresguerres:1995js,tresguerres1995exact,Hehl:1999sb,Garcia:2000yi,Puetzfeld:2001hk,Chen:2009at,Lu:2016bcx,Cembranos:2016gdt,Cembranos:2017pcs,Blagojevic:2017wzf,Blagojevic:2017ssv,Obukhov:2019fti,Bahamonde:2020fnq,Obukhov:2020hlp,Iosifidis:2020gth,Aoki:2020zqm,Bahamonde:2021akc,Bahamonde:2021qjk,delaCruzDombriz:2021nrg,Jimenez-Cano:2022arz,Bahamonde:2022meb,Bahamonde:2022kwg,Yang:2022icz,Bulgakova:2022bsk,Iosifidis:2022xvp,Aoki:2023sum,DeMartino:2023qkl,Mustafa:2023ngp,Yasir:2024nfn,Barker:2024dhb}, where the new degrees of freedom from torsion and nonmetricity significantly impact gravity, reflecting its intrinsic properties~\cite{Hehl:1994ue,ponomarev2017gauge,Cabral:2020fax}.

In these non-Riemannian gravity theories, the form of the Weyl anomaly might also be modified. For example, the form of the anomaly in the Riemann-Cartan case has been discussed by several authors~\cite{Obukhov:1983mm,Goldthorpe:1979ib,Yajima:1985jd,Nieh:1981xk,Buchbinder:1985ym,Gusynin:1988zt,Cognola:1988qw,Gusynin:1990ek,Helayel-Neto:1999acz,Camargo:2022gcw,Paci:2024ohq}, while in the more general case of MAG, particularly regarding the effects of nonmetricity, it has not yet been thoroughly explored, to the best of our knowledge. Our primary objective is to derive a generic form of the Weyl anomaly by considering both the torsion and nonmetricity tensors.

In the generic case of MAG, a new source (or matter content) emerges, associated with the additional degrees of freedom. This quantity is called the hypermomentum tensor, which is defined by taking the variation of a matter action with respect to an affine connection. Depending on the definition of matter coupling, one may have hypermomentum or not. In the minimal coupling scenario, there are no couplings between a scalar field and the connection, leading to a zero hypermomentum~\cite{BeltranJimenez:2020sih}. However, in more general situations, one can have couplings between these quantities, such as by changing the kinetic term from $ \frac{1}{2} g^{\mu\nu}\tilde{\nabla}_\mu \phi \tilde{\nabla}_\nu \phi\equiv \frac{1}{2} g^{\mu\nu}\partial_\mu \phi \partial_\nu \phi$ to $ -\frac{1}{2}g^{\mu\nu}\phi\tilde{\nabla}_\mu \tilde{\nabla}_\nu\phi$~(see for exmple~\cite{Kubota:2020ehu}).

In a similar way, there are several scale transformations in the generic case of MAG since a metric and an affine connection are independent at the action level and can transform independently under a scale transformation. Among them, there are three particularly interesting scale transformations \cite{Iosifidis:2018zwo}. One is the conformal or Weyl transformation, in which only the metric is transformed, and the trace of the energy-momentum tensor vanishes classically if the action is invariant under this transformation. Another is the frame rescaling transformation, in which both the metric and the affine connection are transformed, and the combination of the traces of the energy-momentum tensors and the hypermomentum tensor vanishes classically if the action is invariant under this transformation. Yet another is the projective transformation, in which only the affine connection is transformed, and the trace of the hypermomentum tensor vanishes classically if the action is invariant under this transformation. In particular, the last one is quite special in that only the affine connection is transformed. Therefore, any lesson from a framework of Riemannian geometry does not apply, and it is a non-trivial question whether a new type of trace anomaly appears — that is, whether the trace of the hypermomentum tensor becomes non-zero due to quantum effects under this projective transformation. As far as we know, nobody has yet addressed this question, and this is the second main topic of this paper.

This paper is constructed as follows: In Sec.~\ref{sec:2}, we briefly present the fundamental principles and geometrical descriptions of MAG.
In Sec.~\ref{sec:3}, we review the heat kernel method via the derivation of the trace anomaly for the stress-energy tensor in conformally invariant theories in the framework of Riemannian geometry.
In Sec.~\ref{sec:4}, we go back to the derivation of the asymptotic expansion of the heat kernel and then generalize the expansion coefficients from the Riemannian case and Riemann-Cartan cases to the generic case of MAG. In the first half, we roughly introduce the mathematical tools to construct the heat kernel in the context of the theory for pseudo differential operators. Using these concepts for the heat operator, we follow the algorithm for deriving the Minakshisundaram coefficients, which is known as Seeley's algorithm. In the second part of this section, we review the Riemannian and the Riemann-Cartan cases and then generalize to the MAG case. In this part, we emphasize that the advantage of Seeley's algorithm is that the calculation is systematic and thus easy to modify.
In Sec.~\ref{sec:5}, we apply the generalized coefficients to several concrete theories. First, we consider the trace anomaly in the context of MAG. Since there are independent fundamental variables, the metric and the affine connection, there are three types of scaling transformation: conformal transformation, projective transformation, and frame rescaling. We investigate whether the trace anomaly appears in the scale-invariant theories under these transformations. In particular, it is demonstrated that the trace anomaly emerges in the relationship between the stress-energy tensor and the hypermomentum tensor when the scale-invariant theory is considered under the frame rescaling. Furthermore, we apply our result to the spin $1/2$ case in which fermion couples to the torsion and the nonmetricity via non-minimal coupling.
Finally, in Sec.~\ref{sec:conclusion}, we summarize our main results. 

We will use the mostly negative signature $\eta_{\mu\nu}=\textrm{diag}(+,-,-,-)$ and the natural units $c=\hbar=1$.
For the indices on tensors, we denote spacetime indices and internal indices in Greek and Latin, respectively.
%
%
%%%%%%%%%%%%%%%%%%%%%%%%%%%%%%%%%%%%%%%%%%%%%%%%%%%%%%%
\section{Brief description of Metric-Affine Geometries}\label{sec:2}
We consider a 4-dimensional differentiable manifold equipped with two independent quantities: the metric $g_{\mu\nu}$ and an affine connection $\tilde{\Gamma}^{\lambda}{}_{\mu\nu}$. The latter provides a means to define parallel transport, and in generic MAG geometries, it contains up to 64 degrees of freedom, which can be separated as:
\begin{equation}
\tilde{\Gamma}^{\lambda}\,_{\mu \nu}= \Gamma^{\lambda}\,_{\mu \nu}+ N^{\lambda}\,_{\mu \nu}\,,
\end{equation}
where $\Gamma^{\lambda}\,_{\mu \nu}$ is the Levi-Civita connection and the post-Riemannian terms are described by the so-called distortion tensor $N^{\lambda}\,_{\mu \nu}$, given by
\begin{eqnarray}
    N^{\lambda}\,_{\mu \nu}=K^{\lambda}\,_{\mu \nu}+L^{\lambda}\,_{\mu \nu}\,,
\end{eqnarray}
where the contortion $ K^{\lambda}\,_{\mu \nu}$ and disformation tensors $L^{\lambda}\,_{\mu \nu}$ are defined as:
\begin{align}
    K^{\lambda}\,_{\mu \nu}&=\frac{1}{2}\left(T^{\lambda}\,_{\mu \nu}-T_{\mu}\,^{\lambda}\,_{\nu}-T_{\nu}\,^{\lambda}\,_{\mu}\right)\,,\quad
    L^{\lambda}\,_{\mu \nu}=\frac{1}{2}\left(Q^{\lambda}\,_{\mu \nu}-Q_{\mu}\,^{\lambda}\,_{\nu}-Q_{\nu}\,^{\lambda}\,_{\mu}\right)\,.
\end{align}
In these equations, we have defined the torsion and nonmetricity tensors as
\begin{equation}
    T^{\lambda}\,_{\mu \nu}=2\tilde{\Gamma}^{\lambda}\,_{[\mu \nu]}\,, \quad Q_{\lambda \mu \nu}=\tilde{\nabla}_{\lambda}g_{\mu \nu}\,,
\end{equation}
where $\tilde{\nabla}_\mu$ denotes the covariant derivative related to the general connection acting on an arbitrary tensor $T^{\lambda_{1}...\lambda_{i}}\,_{\rho_{1}...\rho_{j}}$ defined as
\begin{align}
\tilde{\nabla}_\mu T^{\lambda_{1}...\lambda_{i}}\,_{\rho_{1}...\rho_{j}}=&\;\partial_\mu T^{\lambda_{1}...\lambda_{i}}\,_{\rho_{1}...\rho_{j}}+\tilde{\Gamma}^{\lambda_{1}}{}_{\sigma\mu}T^{\sigma...\lambda_{i}}\,_{\rho_{1}...\rho_{j}}+...+\tilde{\Gamma}^{\lambda_{i}}{}_{\sigma\mu_{i}}T^{\lambda_{1}...\sigma}\,_{\rho_{1}...\rho_{j}}\nonumber\\
&-\tilde{\Gamma}^{\sigma}{}_{\rho_{1}\mu}T^{\lambda_{1}...\lambda_{i}}\,_{\sigma...\rho_{j}}-...-\tilde{\Gamma}^{\sigma}{}_{\rho_{j}\mu}T^{\lambda_{1}...\lambda_{i}}\,_{\rho_{1}...\sigma}\,. \label{devG}
\end{align}

Using these defintions we can define the curvature tensor as
\begin{align}\label{totalcurvature}
\tilde{R}^{\lambda}\,_{\rho \mu \nu}&=\partial_{\mu}\tilde{\Gamma}^{\lambda}\,_{\rho \nu}-\partial_{\nu}\tilde{\Gamma}^{\lambda}\,_{\rho \mu}+\tilde{\Gamma}^{\lambda}\,_{\sigma \mu}\tilde{\Gamma}^{\sigma}\,_{\rho \nu}-\tilde{\Gamma}^{\lambda}\,_{\sigma \nu}\tilde{\Gamma}^{\sigma}\,_{\rho \mu}\\
&=R^{\lambda}\,_{\rho\mu\nu}+\nabla_{\mu}N^{\lambda}\,_{\rho \nu}-\nabla_{\nu}N^{\lambda}\,_{\rho \mu}+N^{\lambda}\,_{\sigma \mu}N^{\sigma}\,_{\rho \nu}-N^{\lambda}\,_{\sigma \nu}N^{\sigma}\,_{\rho \mu}\,,
\end{align}
where one notices that the Riemannian curvature tensor $R^{\lambda}\,_{\rho\mu\nu}$ is corrected due to torsion and nonmetricity.

For our study, it would be useful to decompose torsion and nonmetricity in the pseudo-orthogonal group as:
\begin{align}\label{irreducibletorsion}
T^{\lambda}\,_{\mu \nu}&=\frac{1}{3}\left(\delta^{\lambda}\,_{\nu}T_{\mu}-\delta^{\lambda}\,_{\mu}T_{\nu}\right)+\frac{1}{6}\,\varepsilon^{\lambda}\,_{\rho\mu\nu}S^{\rho}+t^{\lambda}\,_{\mu \nu}\,,\\
Q_{\lambda\mu\nu}&=g_{\mu\nu}W_{\lambda}+{\nearrow\!\!\!\!\!\!\!Q}_{\lambda\mu\nu}=g_{\mu\nu}W_{\lambda}+g_{\lambda(\mu}\Lambda_{\nu)}-\frac{1}{4}g_{\mu\nu}\Lambda_{\lambda}+\frac{1}{3}\varepsilon_{\lambda\rho\sigma(\mu}\Omega_{\nu)}\,^{\rho\sigma}+q_{\lambda\mu\nu}\,,\label{Qdecomposition}
\end{align}
where ${\nearrow\!\!\!\!\!\!\!Q}_{\lambda\mu\nu}$ is the so-called traceless part of nonmetricity. For the torsion sector, we have introduced one vector $T_\mu$, one pseudo-vector $S_\mu$ and one totally traceless tensor $t_{\lambda\mu\nu}$ (satisying $t_{[\lambda\mu\nu]}=0$). These quantities are defined as
\begin{align}\label{Tdec1}
T_{\mu}&=T^{\nu}\,_{\mu\nu}\,,\\
S_{\mu}&=\varepsilon_{\mu\lambda\rho\nu}T^{\lambda\rho\nu}\,,\\
t_{\lambda\mu\nu}&=T_{\lambda\mu\nu}-\frac{2}{3}g_{\lambda[\nu}T_{\mu]}-\frac{1}{6}\,\varepsilon_{\lambda\rho\mu\nu}S^{\rho}\,.\label{Tdec3}
\end{align}
On the other hand, for the nonmetricity sector, we introduce two vectors: $W_{\mu}$ (the so-called Weyl part of nonmetricity) and $ \Lambda_{\mu}$, along with one totally traceless and pseudo-traceless tensor $\Omega_{\lambda}{}^{\mu\nu}$ (also satisfying $\Omega_{[\lambda\mu\nu]}=0$), and finally, a totally symmetric tensor $q_{\lambda\mu\nu}$, defined as:
\begin{align}
W_{\mu}&=\frac{1}{4}\,Q_{\mu\nu}\,^{\nu}\,,\label{weyl}\\
     \Lambda_{\mu}&=\frac{4}{9}\left(Q^{\nu}\,_{\mu\nu}-W_{\mu}\right)\,,\\
    \Omega_{\lambda}\,^{\mu\nu}&=-\,\left[\varepsilon^{\mu\nu\rho\sigma}Q_{\rho\sigma\lambda}+\varepsilon^{\mu\nu\rho}\,_{\lambda}\left(\frac{3}{4}\Lambda_{\rho}-W_{\rho}\right)\right]\,,\\
    q_{\lambda\mu\nu}&=Q_{(\lambda\mu\nu)}-g_{(\mu\nu}W_{\lambda)}-\frac{3}{4}g_{(\mu\nu}\Lambda_{\lambda)}\,.\label{qtensor}
\end{align}
where $\varepsilon_{\mu\nu\rho\sigma}=\sqrt{-g}\epsilon_{\mu\nu\rho\sigma}$ is the 4-dimensional Levi-Civita tensor and $\epsilon_{\mu\nu\rho\sigma}$ the 4-dimensional Levi-Civita symbol.

There are different types of geometries that can be obtained from the generic MAG geometry case. For example, the simplest of them is when torsion and nonmetricity vanish, providing us with Riemannian geometry. Other important examples include when nonmetricity vanishes, giving rise to the so-called Riemann-Cartan geometries, or when only nonmetricity is described solely by its Weyl part, labeled as Weyl-Cartan geometries.

We can then formulate theories of gravity in generic MAG geometries as:
\begin{eqnarray}
    S(g,\tilde{\Gamma})=\int d^4 x \sqrt{-g}\Big[L_{\rm g}(g,\tilde{\Gamma})+L_{\rm m}(g,\tilde{\Gamma})\Big]\,,
\end{eqnarray}
with $L_{\rm g}$ denoting the gravitational sector and $L_{\rm m}$ the matter content. Since the matter Lagrangian depends on the metric and the connection, we can define two different matter quantities. First, we can define the energy-momentum tensor  as
\begin{eqnarray}
     T_{\mu\nu} =- \frac{2}{\sqrt{-g}}\frac{\delta (\sqrt{-g}L_{\rm m})}{\delta g^{\mu\nu}}\,.
\end{eqnarray}
Further, we can define a new source labeled as hypermomentum defined as
\begin{eqnarray}
  {\Delta_\lambda}^{\mu\nu}=  {-\frac{2}{\sqrt{-g}}} \frac{ \delta(\sqrt{-g}L_{\rm m})}{\delta \tilde{\Gamma}^\lambda{}_{\mu\nu}}\,.
\end{eqnarray}
This quantity can be further split into three pieces:
\begin{eqnarray}\label{hypermomentum}
 \Delta_{\mu\nu\lambda}={}^{(\rm s)}\Delta_{[\mu\nu]\lambda}+\frac{1}{4}g_{\mu\nu}{}^{(\rm d)}\Delta_{\lambda}+{}^{(\rm sh)}{\nearrow\!\!\!\!\!\!\!\Delta}_{(\mu\nu)\lambda}\,.
\end{eqnarray}
Here, the first part is related to intrinsic spin (which is the source of torsion), the second part is related to intrinsic dilations (which is the source of the trace part of nonmetricity) with ${}^{(\rm d)}\Delta_{\mu}=\Delta^{\nu}{}_{\nu\mu}$, and the last part is related to intrinsic shears (which provide a source for the traceless part of nonmetricity).

%%%%%%%%%%%%%%%%%%%%%%%%%%%%%%%%%%%%%%%%%%%%%%%%%%%%%%%
\section{Heat Kernel method}\label{sec:3}
The heat kernel method is a useful prescription for evaluating one-loop effective actions and their variations. One interesting quantum effect that can be analyzed using this method is the trace anomaly, which we will focus on hereafter.
It is then worthwhile to review its broad outlines in the context of GR before discussing anomalies in MAG.
This section briefly reviews the heat kernel method and the Weyl anomaly in the GR case.
Some mathematical concepts and definitions introduced in this section will be used in the following sections when considering the generalization of trace anomalies to MAG.
\subsection{One-loop effective action}
Let us consider the simplest case with \(N\)-dimensional real scalar fields \(\vec{\phi} = (\phi_1, \phi_2, \cdots, \phi_N)^T\) and its action \(S_{\rm m}\) in the presence of a gravitational field, given by:
\begin{align}
S_{\rm m} = \frac{1}{2} \int \sqrt{-g} \, d^4x \, \vec{\phi}^\dagger \, \mathcal{O} \, \vec{\phi} \,, \label{action-Sm}
\end{align}
where  \(\mathcal{O}\) is a second-order differential 
 operator acting on the fields that we will assume to be of the form
\begin{align}
\mathcal{O}=
    -\qty(
        g^{\mu\nu}(x) \tilde{\nabla}_\mu \tilde{\nabla}_\nu \mathbbm{1}_N
        + \mathbb{A}^\mu(x) \tilde{\nabla}_\mu
        + \mathbb{X}(x)
    )\,.\label{eqO}
\end{align}
Here, we introduce the identity matrix $\mathbbm{1}_N \in M_{N \times N}(\mathbb{R})$ and coefficients on covariant derivatives.
The coefficients $\mathbb{A}^\mu(x)$ and $\mathbb{X}(x)$ are $N \times N$ matrix and, in general, they are matrix-valued functions depending on the coordinate $x$.
In what follows, we will always denote matrix-valued quantities by a blackboard bold character.
For instance, we denote the matrix with a Lorentz index like $\mathbb{A}^{\mu}=\mathbb{A}^{\mu}(x)$.

Now let us impose that, under the scale transformation on the metric $g_{\mu\nu}\longrightarrow e^{2k(x)}g_{\mu\nu}$ with an arbitrary function $k(x)$, the set of fields $\vec{\phi}(x)$ transforms as
\begin{align}
    \vec{\phi}(x) \longrightarrow e^{-wk(x)} \vec{\phi}(x) \label{weyl-trans-field}\,,
\end{align}
where $w$ is known as the conformal weight. This weight is determined such that we can make the kinetic term invariant under this transformation.
For instance, a scalar field with the two-derivative kinetic term has $w=1$.
With this transformation law \eqref{weyl-trans-field}, the operator transforms as:
\begin{align}
    \mathcal{O} \longrightarrow e^{-(4-w)k(x)} \mathcal{O} e^{wk(x)}
    \label{weyl-trans-op}\,,
\end{align}
leaving the matter action invariant. Then, classically, the energy-momentum tensor will be traceless. Nevertheless, this property will be broken after considering a quantum perspective.
Such anomaly is labelled as the Weyl anomaly. In what follows, let us consider the evaluation of this anomaly by the effective action.
When we consider the path integral and the effective action, we usually introduce the eigenvalues $\lambda_n$ and eigenfunctions $\vec{\varphi}_n$ of $\mathcal{O}$ which satisfy
\begin{align}
    \mathcal{O} \vec{\varphi}_n(x) = \lambda_n \vec{\varphi}_n(x)
\end{align}
with the orthonormal condition and the normalization in the presence of the curved spacetime:
\begin{align}
    \int \sqrt{-g} d^4x \vec{\varphi}^\dagger_n(x) \vec{\varphi}_m(x) = \delta_{nm}
\qquad
    \sum_n \vec{\varphi}^\dagger_n(x) \vec{\varphi}_n(y) = \frac{\delta(x-y)}{\sqrt[4]{-g(x)}\sqrt[4]{-g(y)}}\,.
\end{align}
By assuming that the set of these eigenfunctions $\{\vec{\varphi}_n(x)\}$ is a complete orthonormal system, we can expand the set of fields $\vec{\phi}$ with dimensionless coefficients $a_n$ on the eigenfunction $\vec{\varphi}_n(x)$ as follows%
\begin{align}
    \vec{\phi}(x) = \sum_n a_n \vec{\varphi}_n(x)\,.
\end{align}
Under this expansion, the integral measure $[\mathcal{D}\vec{\phi}]$ in the path integral is represented by the product of the measure $da_n$ as the following equation:
\begin{align}
    [\mathcal{D}\vec{\phi}] = \prod_n \frac{d a_n}{\sqrt{2\pi}}\,,
\end{align}
where we have introduced a normalization factor $(2\pi)^{-1/2}$.
After performing the path integral and a Legendre transformation, the one-loop effective action $\Gamma[g_{\mu\nu}]$ becomes
\begin{align}
\Gamma[g_{\mu\nu}]=
    S[g_{\mu\nu}]
    + \frac{1}{2} \log \det \mathcal{O}\,,
\end{align}
where $\det \mathcal{O}$ denotes the functional determinant of the operator $\mathcal{O}$  defined as 
\begin{align}
    \det \mathcal{O} = \prod_n \lambda_n\,,
\end{align}
which is the product of all its eigenvalues.
\subsection{The heat kernel method}
Since the product of all the eigenvalues is not a finite quantity, we need to regularize it. To do this, one usually introduce spectral zeta function
$\zeta(\mathcal{O},s)$ defined by
\begin{align}
    \zeta(\mathcal{O},s) = \sum_n \qty(\dfrac{1}{\lambda_n})^s\,,
\end{align}
where the sum converges for sufficiently large real $s$ and then becomes an analytical function \cite{Hawking:1976ja}.
When the differential operator $\mathcal{O}$ is a Laplacian on a compact Riemannian manifold, this zeta function is called as Minakshisundaram-Pleijel zeta function~\cite{Gilkey:1975iq,Obukhov:1983mm,Minakshisundaram:1949xg}.
The spectral zeta function gives the representation for the logarithm of the functional determinant:
\begin{align}
    \log\det\mathcal{O} = -\left.\frac{\partial\zeta(\mathcal{O},s)}{\partial{s}}\right|_{s=0}\,.
\end{align}
To be precise, this equation defines the functional determinant of the operator.
The heat kernel method gives the representation of this zeta function $\zeta(\mathcal{O},s)$.
At first, we consider the heat operator, defined by $e^{-\mathcal{O}\tau}$ with the parameter $\tau$.
The heat kernel
$\mathbb{K}(e^{-\mathcal{O}\tau};x,y)$ is defined by the following equation:
\begin{align}
    e^{-\mathcal{O}\tau} \vec{u}(x) = \int \sqrt{-g} dy \mathbb{K}(e^{-\mathcal{O}\tau};x,y) \vec{u}(y)\,,
\end{align}
with an arbitrary $N$-component function $\vec{u}(x)$.
Under the basis of the eigenfunction $\vec{\varphi}_n(x)$, we can find the representation of this heat kernel by
\begin{align}
\mathbb{K}(e^{-\mathcal{O}\tau};x,y)
= \sum_n e^{-\lambda_n \tau} \vec{\varphi}_n(x) \vec{\varphi}^\dagger_n(y)\,.
\label{heat-kernel-rep-eigen}
\end{align}
Note that this heat kernel $\mathbb{K}(e^{-\mathcal{O}\tau};x,y)$ is given as a matrix-valued function.
From this representation, the trace of the diagonal part in the heat kernel is represented as
\begin{align}
\Tr e^{-\mathcal{O}\tau}
= \int \sqrt{-g} dx \tr \mathbb{K}(e^{-\mathcal{O}\tau};x,x)
= \sum_n e^{-\lambda_n \tau}\,.
\label{tr-kernel}
\end{align}
The trace operator $\tr$ is a usual trace of the matrix and the trace operator $\Tr$ is the trace for all degrees of freedom, which is defined by the following equation.
\begin{align}
\Tr\qty[\cdots]
= \int dx \sqrt{-g} \sum_n \vec{\varphi}_n^\dagger(x) \qty[\cdots] \vec{\varphi}_n(x)\,.
\end{align}
For a positive number $\lambda$, we can use the identity for the Melin transformation of $\tau^{s-1}$:
\begin{align}
    \frac{1}{\lambda^s} = \int\frac{d\tau}{\Gamma(s)} \tau^{s-1} e^{-\lambda \tau}\,.
\end{align}
By applying this identity for each eigenvalue $\lambda_n$, we can find the representation for the zeta function by the trace of the heat kernel
\begin{align}
\zeta(\mathcal{O},s)
= \int\frac{d\tau}{\Gamma(s)} \tau^{s-1} \Tr e^{-\mathcal{O}\tau}\,.
\end{align}
Further, the diagonal part of the heat kernel $\mathbb{K}(e^{-\mathcal{O}\tau};x,x)$ has the following asymptotic expansion as $\tau\to+0$:
\begin{align}
\tr \mathbb{K}(e^{-\mathcal{O}\tau};x,x)
\sim \frac{\sqrt{-g}}{(4\pi\tau)^{d/2}} \sum_n c_n(x) \tau^n\,,
\label{HMDS-expansion}
\end{align}
where $d$ is the dimension of the manifold.
This asymptotic expansion is called as the Minackshisundaram-Pleijel expansion and the expansion coefficients are called as the Hadamard-Minakshisundaram-DeWitt-Seeley (HMDS) coefficients~\cite{DeWitt:1964mxt,Minakshisundaram:1949xg,seeley1967complex}.
The evaluation of the anomaly on the curved spacetime strongly relates to these coefficients.

\subsection{The Weyl anomaly in Riemannian geometry}

When we evaluate the variation of the effective action, the proper time representation is useful. By using this we have
\begin{align}
S_{\text{eff}}[g_{\mu\nu}]
    = S[g_{\mu\nu}]
        - \frac{1}{2} \int^\infty_{1/\Lambda^2} \frac{ds}{s} \Tr e^{-s\mathcal{O}}\,,
\end{align}
where $\Lambda$ is a cut-off scale of the UV-divergence. From this representation, we can define the variation of the effective action as follows:
\begin{align}
\delta{ {S_{\text{eff}}}}[g_{\mu\nu}]
=   \frac{1}{2} \int^\infty_{1/\Lambda^2} ds \Tr \qty(e^{-s\mathcal{O}}\delta\mathcal{O})
=   \int d^4x \frac{\sqrt{-g}}{(4\pi)^2}
    \qty[
        c_0(x) \Lambda^4
        + c_1(x) \Lambda^2
        + c_2(x) 
        + c_3(x) \qty(\frac{1}{\Lambda^2})
        + \cdots
    ] \omega(x)\,.
\label{variation-via-heat-kernel}
\end{align}
In the above equation, the first two terms (that are proportional to $\Lambda^2$ and $\Lambda^4$) must be removed by a suitable renormalization scheme. After doing so, we can find a finite result in the limit $\Lambda\to\infty$.
Under the transformation law \eqref{weyl-trans-op}, the infinitesimal transformation of the operator $\mathcal{O}$ is given by
\begin{align}
\delta\mathcal{O}
    = -(4-w) \omega(x) \mathcal{O} + w \mathcal{O} \omega(x) \label{variation-operator} \,.
\end{align}
By substituting this expression \eqref{variation-operator} into Eq. \eqref{variation-via-heat-kernel}, we find
\begin{align}
\delta{ {S_{\text{eff}}}}
=   (2-w) \int d^4 x \Tr \qty(e^{-s\mathcal{O}}\omega)\,.
\label{vary-gamma}
\end{align}
On the other hand, the variation of the effective action $\delta{ {S_{\text{eff}}}}$ gives the expectation value of the energy-momentum tensor:
\begin{align}
\delta{ {S_{\text{eff}}}}
    = \int d^4x \sqrt{-g} \expval{T_\mu^{\phantom{\mu}\mu}} \omega(x)\,.
\end{align}
By comparing this variation to \eqref{vary-gamma} with \eqref{variation-via-heat-kernel}, we can evaluate the expectation value of the trace in the energy-momentum tensor.
Thus, the final result for varying the effective action gives
\begin{align}
\delta{ {S_{\text{eff}}}} = \int d^4x \frac{\sqrt{-g}}{(4\pi)^2} c_2(x)
\qquad
\expval{T_\mu^{\phantom{\mu}\mu}} = \frac{c_2(x)}{(4\pi)^2}\,.
\end{align}
Finally, we can evaluate the variation of the effective action and the trace anomaly for the energy-momentum tensor by using the asymptotic expansion of the heat kernel \eqref{HMDS-expansion}.
In this expansion, the expansion coefficients $c_n(x)$ are strongly dependent on the geometry of spacetime.
So, when we consider the heat kernel method in the non-Riemannian spacetime, we have to modify these coefficients based on the geometry of the spacetime in which we apply this method. Furthermore, in the presence of a hypermomentum, another term would appear in the above expressions.
In the following sections, we will go back to the derivations for these expansion coefficients.

\section{Seeley's algorithm and HMDS coefficients in Metric-Affine gravity}\label{sec:4}

In the derivation for Seeley-DeWitt expansion, there are several methods \cite{DeWitt:1964mxt, Gilkey:1975iq}.
In this paper, we consider the approach based on the theories of (pseudo) differential operators.
The Riemannian case was found by Gilkey in~\cite{Gilkey:1975iq}. Then, Obukhov generalised such a result for the Riemann-Cartan case, which includes an additional torsion degree of freedom to the Riemannian case \cite{Obukhov:1983mm}. Both studies are mainly based on the work of Seeley \cite{seeley1967complex}.
In this section, we introduce Seeley's algorithm and then apply this method to the Riemannian case and generic MAG.
To emphasize that the differential operator in the following discussion is generic, we will denote it by $D$.

\subsection{Seeley's algorithm}

\subsubsection{The resolvent and symbol}
Let us first introduce the mathematical concepts related to the theory of pseudo-differential operators.
Consider the (pseudo) differential operator $D$ which acts on arbitrary functions.
Let us denote $u(x)$ as an arbitrary function on $\mathbb{R}^n$ and its Fourier transformation as $\hat{u}(\xi)$.
\begin{align}
\begin{aligned}
\hat{u}(\xi) = \int \frac{d^n\xi}{(2\pi)^n} e^{- i x^\mu \xi_\mu} u(\xi)\,.
\qquad
u(x) = \int {d^n x} e^{i x^\mu \xi_\mu} \hat{u}(\xi)\,,
\end{aligned}
\label{eq:fourier-trans-for-function-U}
\end{align}
Here we denote the Fourier variables by the covector $\xi_\mu$.
The action of the differential operator $D$ on $u(x)$ will be represented by using this Fourier transformation pair
\begin{align}
Du(x)=
    \int \frac{d^n\xi}{(2\pi)^n}
        ~ e^{i x^\mu \xi_\mu}
        ~ \sigma(D)(x,\xi)
        ~ \hat{u}(\xi)\,,
    \label{def-symbol}
\end{align}
where $\sigma(D)(x,\xi)$ is a polynomial in $x^\mu$ and $\xi_\mu$ that is usually labelled as the symbol of the differential operator $D$.
Between two given symbols $p_1(x,\xi)=\sigma(D_1)(x,\xi)$ and $p_2(x,\xi)=\sigma(D_2)(x,\xi)$, the symbol of the product $\sigma(D_1 \cdot D_2)(x,\xi)$ is given by the following equation \cite{Kohn1965}:
\begin{align}
\sigma(D_1 \cdot D_2)(x,\xi)
= \sum_{l=0}^\infty\frac{(-1)^l}{l!}
    \qty(
        V^{\alpha_1}_\xi V^{\alpha_2}_\xi \cdots V^{\alpha_l}_\xi p_1(x,\xi)
    )
    \qty(
        \partial_{\alpha_1}^x \partial_{\alpha_2}^x \cdots \partial_{\alpha_l}^x p_2(x,\xi)
    )
\label{product-law}
\end{align}
with $V^\alpha_\xi := \partial/\partial\xi_\alpha$.
Let us now introduced the so-called kernel, which is constructed from the differential operator $D$, $\mathbb{K}(D;x,y)$, and is defined as:
\begin{align}
Du(x) =  \int d^n y ~ \mathbb{K}(D;x,y) ~ u(y) \,.\label{def-kernel}
\end{align}
The heat kernel introduced in the previous section has a similar definition. Now, for a differential operator $D$, there is an important relation between the symbol $\sigma(D)(x,\xi)$ and the kernel $\mathbb{K}(D;x,y)$.
By substituting an equation of the Fourier transformation for $u(x)$ \eqref{eq:fourier-trans-for-function-U} into \eqref{def-kernel} and comparing it with \eqref{def-symbol}, we obtain 
\begin{align}
\mathbb{K}(D;x,y)
    = \int\frac{d^n\xi}{(2\pi)^n} ~ \sigma(D)(x,\xi) e^{i(x^\mu-y^\mu)\xi_\mu}\,.
\end{align}
This equation implies that the trace part ($y\to x$) of the kernel has the following integral representation formula:
\begin{align}
\mathbb{K}(D;x,x)
    = \int\frac{d^n\xi}{(2\pi)^n} ~ \sigma(D)(x,\xi) \label{int_rep-trace_of_kernel}\,.
\end{align}
This is an important property connecting the kernel and the symbol. Note that $\mathbb{K}(D;x,x)$ corresponds to the diagonal part of the heat kernel.
\subsubsection{The heat operator and kernel}
The heat operator of $D$ has the following integral representation
\begin{align}
e^{-\tau D}
    = \frac{i}{2\pi} \int_C d\lambda ~
        e^{-\tau \lambda} ~ \qty(D-\lambda)^{-1}
\qquad \text{for}~~ \tau > 0\,.
\label{int_rep-heat_kernel}
\end{align}
The integration contour \( C \) is defined so that its interior encompasses the spectrum of \( D \) \cite{Gilkey:1975iq, Obukhov:1983mm}.
.
In this equation, the operator $\qty(D-\lambda)^{-1}$ is called the resolvent. The core of the heat kernel's asymptotic expansion comes from this resolvent's asymptotic expansion.
Let us apply this representation to the definition of symbol \eqref{def-symbol} with $P=\exp(-\tau D)$:
\begin{align}
\sigma\qty(e^{-\tau D})
        = \frac{i}{2\pi} \int_C d\lambda ~
            e^{-\tau \lambda} ~ \sigma\qty(\qty(D-\lambda)^{-1})\,.
\label{symbol-heat-op}
\end{align}
With this quantity, we can construct the asymptotic expansion of the symbol \(\sigma((\lambda - D)^{-1})(x, \xi)\) for the heat operator \(P = \exp(-\tau D)\). This means that, through \eqref{int_rep-trace_of_kernel}, we can obtain the trace \(\Tr \mathbb{K}(P; x, x)\) as an asymptotic expansion, known as the Seeley-DeWitt expansion.

\subsubsection{The asymptotic expansion for the symbol of resolvent}
In general, we can consider the differential operator $D$ acting on fields with multiple $N$-components where $N$ is defined as a positive integer number.
For instance, the simplest example is the multiple scalar field case.
Also, in the Dirac theory, the corresponding differential operator will be identified with the square of the Dirac operator $\gamma^\mu\qty(\partial_\mu-m)$.
As in~\eqref{eqO}, let us then consider the following second-order differential operator $D$ as:
\begin{align}
D &= -\qty(
        g^{\mu\nu} \mathbbm{1}_N \tilde{\nabla}_\mu \tilde{\nabla}_\nu
        + \mathbb{A}^\mu \tilde{\nabla}_\mu
        + \mathbb{X}
    )\,,\label{gen-cov-deriv}\\
    &= -\qty(
        g^{\mu\nu} \partial_\mu \partial_\nu \mathbbm{1}_{N}
        + \mathbb{H}^\mu \partial_\mu
        + \mathbb{E}
    )\,,
\label{expand_gen-cov-deriv}
\end{align}
where in the last step we have expanded the generalised covariant derivative that was defined in~\eqref{devG} and redefined the coefficient matrix $\mathbb{H}^\mu=\mathbb{H}^\mu(x)$ and $\mathbb{E}=\mathbb{E}(x)$ by substituting the definition of the covariant derivative $\tilde{\nabla}_\mu=\partial_\mu+\tilde{\omega}_\mu+\tilde{\Gamma}_\mu$ with the partial derivative $\partial_\mu$, the spin connection $\tilde{\omega}_\mu$, and the affine connection $\Tilde{\Gamma}^\lambda{}_{\mu\nu}$.
For instance, the covariant derivative $\tilde{\nabla}_\mu$ act on a scalar field $\phi$ and a fermion field $\psi$ as follows
\begin{align}
\tilde{\nabla}_\mu \phi
&=  \partial_\mu \phi
\label{eq:cov-on-scalar}
    \\
\tilde{\nabla}_\mu \psi
&=  
    \qty{
        \partial_\mu
        + \frac{1}{8} \tilde{\omega}_\mu{}^{ab} \qty[\gamma_a,\gamma_b]
    } \psi\,,
\label{eq:cov-on-fermion-assumed}
\end{align}
where $\tilde{\omega}_\mu{}^{ab}$ is the component of the spin connection $\tilde{\omega}_\mu$.
We should note that the group where the gauge approach of MAG is constructed is the so-called General Linear group $GL(4,R)$ that is an infinite dimensional gauge group~\cite{Hehl:1994ue}. Therefore, any coupling between a spin $1/2$ and a general spin connection is somehow artificial or introduced by hand. However, we assumed the above form as one possible coupling that can exist in this framework and only introduce interactions between the antisymmetric part of the connection and the spin $1/2$. This means that we assumed that the hypermomentum associated with this field would not contain any shear current.
As we will notice later, the coefficient matrices will depend on an affine connection for the curved space-time.
In the component representation, this differential operator $D$ can be written as
\begin{align}
D_{ij} =
    -\qty(
        \delta_{ij} g^{\mu\nu} \partial_\mu \partial_\nu
        + \qty(\mathbb{H}^\mu)_{ij} \partial_\mu
        + \qty(\mathbb{E})_{ij}
    )\,,
\label{expand_gen-cov-deriv_coord}
\end{align}
where the indices $i,\,j$ run from $1,2,...,N$.
This operator acts on the $N$-component field $\Phi$ as
\begin{align}
    \qty(D\Phi)_i = D_{ij} \Phi_j\,,
\end{align}
where $\Phi_i$ is the $i$-th component of the field $\Phi$.
Now we will asymptotically construct the resolvent. To do this, we consider the symbol $\sigma(B)=\mathbbm{b}(x,\xi,\lambda)$ decomposed as
\begin{align}
\sigma\qty(B)
=
    \mathbbm{b}(x,\xi,\lambda)
\sim
    \mathbbm{b}_0(x,\xi,\lambda)
    + \mathbbm{b}_1(x,\xi,\lambda)
    + \cdots
    + \mathbbm{b}_i(x,\xi,\lambda)
    + \cdots
\label{asymp-symbol-resolvent}
\end{align}
where each symbol $\mathbbm{b}_i(x,\xi,\lambda)$ is dependent only on the coordinate $x^\mu$. Here, we have defined a covector $\xi_\mu$ and a complex parameter $\lambda$. Next, we assume that the symbol $\mathbbm{b}_i$ has a homogeneous order $-2-i$ with respect to the covector $\xi_\mu$.
Thus, we require that the symbol $B$ must obey the following relation
\begin{align}
    \sigma\qty(B(D-\lambda)) = I\,. \label{inv-resolvent}
\end{align}

Based on this relation, we can inductively construct the asymptotic expansion of the symbol of resolvent by looking at the homogeneous order.
From the multiple law for the symbols \eqref{product-law}, $\sigma\qty(B\qty(D-\lambda))$ acquires the following form:
\begin{align}
\sigma(B\qty(D-\lambda))=
    \sum_{p=0}^\infty
        \frac{(-i)^p}{p!}
            \qty(
                {V_\xi}^{\alpha_1}
                \cdots
                {V_\xi}^{\alpha_p}
                \mathbbm{b}
            )
            \qty(
                {\partial^x}_{\alpha_1}
                \cdots
                {\partial^x}_{\alpha_p}
                \sigma(D-\lambda)
            )\,.
\end{align}
By decomposing this sum with respect to the homogeneous order in the covector \(\xi_\mu\), we obtain the following relations:
\begin{align}
\mathbbm{b}_0&=
    \qty(\mathbbm{a}^2-\lambda\mathbbm{1}_N)^{-1}
    = \qty(g^{\mu\nu} \xi_\mu \xi_\nu-\lambda)^{-1} \mathbbm{1}_N\,, \\
\mathbbm{b}_n&=
    - \mathbbm{b}_0 \sum_{j=0}^{n-1} \sum_{p=0}^\infty
        \frac{(-i)^p}{p!}
            \qty(
                {V_\xi}^{\alpha_1}
                \cdots
                {V_\xi}^{\alpha_p}
                \mathbbm{b}_j
            )
            \qty(
                {\partial^x}_{\alpha_1}
                \cdots
                {\partial^x}_{\alpha_p}
                \mathbbm{a}^{p+2+j-n}
            )\,,
\label{recurrence-bi}
\end{align}
where each $\mathbbm{a}_i$ is defined by the following equations
\begin{align}
    \sigma(D) = \mathbbm{a}^2 + \mathbbm{a}^1 + \mathbbm{a}^0\,,
    \qquad
    \mathbbm{a}^2 = g^{\mu\nu} \xi_\mu \xi_\nu \mathbbm{1}_N\,,
    \qquad
    \mathbbm{a}^1 = -i \mathbb{H}^\mu \xi_\mu\,,
    \qquad
    \mathbbm{a}^0 = -\mathbb{E}\,.
\end{align}
Now, we can compute $\mathbbm{b}_n$ by substituting $\mathbbm{b}_{i < n}$, i.e., $\mathbbm{b}_0,\,\mathbbm{b}_1,\,\mathbbm{b}_2,\ldots,\mathbbm{b}_{n-1}$. into \eqref{recurrence-bi}.
For instance, the symbol $\mathbbm{b}_3$ is determined from the symbol $\mathbbm{b}_0$, $\mathbbm{b}_1$, and $\mathbbm{b}_2$.
Then, our aim is to obtain the symbols $\mathbbm{b}_n$ and their coefficients $\mathbbm{b}_{(n)}^{\alpha_1 \alpha_2 \ldots \alpha_p}(x)$ as expressed as:
\begin{align}
\mathbbm{b}_n=
    \sum_{p=0}^\infty
    \qty(\mathbbm{b}_0)^{1+(n+p)/2}~
    \mathbbm{b}_{(n)}^{\alpha_1 \alpha_2 \ldots \alpha_p}(x)~
    \xi_{\alpha_1} \xi_{\alpha_2} \ldots \xi_{\alpha_p}\,.
\label{struct-bn}
\end{align}
This structure is determined from the homogeneous order with respect to the covector $\xi$.
Since these coefficients are cumbersome, we provide them explicitly in the Appendix \ref{app:2_result-of-Seeley}.

\subsubsection{Asymptotic expansion for heat kernel} \label{sssec:asymp-expansion}
Now we are ready to obtain the asymptotic expansion of the heat kernel.
By substituting \eqref{struct-bn} into \eqref{symbol-heat-op}, we obtain the asymptotic form of the symbol for the heat operator
\begin{align}
\sigma\qty(\exp\qty(-D\tau))
=   \sum_{n=0}^\infty \tau^{n/2}
    \qty[
        \sum_{p=0}^\infty
        \qty{\qty(\frac{n+p}{2})!}^{-1} \cdot
        \mathbbm{b}_{(n)}^{\alpha_1 \alpha_2 \ldots \alpha_p}(x) \cdot
        \qty{
            \tau^{p/2} e^{-g^{\mu\nu} \xi_\mu \xi_\nu} \xi_{\alpha_1 \alpha_2 \ldots \alpha_p}
        }
    ]\,.
\end{align}
In this equation, the $\xi$-dependence appears only in the second curly bracket.
This symbol gives the trace of the heat kernel via \eqref{int_rep-trace_of_kernel}:
\begin{align}
\mathbb{K}(\exp\qty(-\tau D);x,x)
=  \sum_{n=0}^\infty \tau^{n/2}
    \qty[
        \sum_{p=0}^\infty
        \qty{\qty(\frac{n+p}{2})!}^{-1} \cdot
        \mathbbm{b}_{(n)}^{\alpha_1 \alpha_2 \ldots \alpha_p}(x) \cdot
        \qty{
            \tau^{-d/2}
            \int_{\mathbb{R}^d} \frac{d\xi}{(2\pi)^d}
            e^{-g^{\mu\nu} \xi_\mu \xi_\nu} \xi_{\alpha_1} \xi_{\alpha_2} \ldots \xi_{\alpha_p}
        }
    ]\,,
\label{trace-of-heat-op_xi}
\end{align}
where for simplicity, we denote the above integral using the following notation
\begin{align}
\int_{\mathbb{R}^d} \frac{d\xi}{(2\pi)^d}
    \xi_{\alpha_1} \xi_{\alpha_2} \ldots \xi_{\alpha_p}
    e^{-g^{\mu\nu} \xi_\mu \xi_\nu}
= \frac{\sqrt{-g}}{(4\pi)^{d/2}} X_{\alpha_1 \alpha_2 \ldots \alpha_p}
\label{int-X}\,.
\end{align}
The quantity \(X_{\alpha_1 \alpha_2 \ldots \alpha_p}\) is constructed from the product of metric tensors \cite{Obukhov:1983mm, Vassilevich:2003xt}. We explicitly provide the first terms in the Appendix \ref{app:1_tensor-X}.
Using this notation, the $\xi$-integrals in \eqref{trace-of-heat-op_xi} are replaced with functions that depend on the coordinates. Then, Eq.~\eqref{trace-of-heat-op_xi} gives us the asymptotic expansion of the heat kernel $\tau$
\begin{align}
\mathbb{K}(\exp\qty(-\tau D);x,x)
=  \frac{\sqrt{-g}}{(4\pi\tau)^{d/2}}
    \sum_{\substack{n=0\\n:\mathrm{even}}}^\infty
    \qty[
        \qty{
            \sum_{p=0}^\infty
            \frac{
                \mathbbm{b}_{(n)}^{\alpha_1 \alpha_2 \ldots \alpha_p}
                X_{\alpha_1 \alpha_2 \ldots \alpha_p}
            }{\qty(\frac{n+p}{2})!}
        }
        \tau^{n/2}
    ]\,.
\label{heat-coeff_symbol}
\end{align}
Therefore, we find the following series in terms of $\tau$:
\begin{align}
\mathbb{K}(\exp\qty(-\tau D);x,x) =
    \frac{\sqrt{-g}}{(4\pi\tau)^{d/2}} \qty[ \mathbb{K}_0 + \tau \mathbb{K}_1 + \tau^2 \mathbb{K}_2 + \cdots ]
\qquad
\mathbb{K}_n = 
    \sum_{p=0}^{\infty}
        \frac{X_{\alpha_1 \alpha_2 \ldots \alpha_p}}{\qty(\frac{2n+p}{2})!}
        \mathbbm{b}_{(2n)}^{\alpha_1 \alpha_2 \ldots \alpha_p}\,,
\label{eq:expansion-of-heat-kernel-matrix}
\end{align}
In the standard notation, the trace of the heat operator is expressed as
\begin{align}
\tr \mathbb{K}(\exp\qty(-\tau D);x,x)
=  \frac{\sqrt{-g}}{(4\pi\tau)^{d/2}}
    \qty[
        c_0(x)
        + c_1(x) \tau
        + c_2(x) \tau^2
        + \cdots
        + c_k(x) \tau^k
        + \cdots
    ]\,.
\label{heat-coeff_a}
\end{align}
where $c_0=1$.
By comparing the above equation with \eqref{heat-coeff_symbol}, we obtain the expression for each coefficient $c_n$:
\begin{align}
c_{n}(x)
=   \sum_{p=0}^\infty
    \frac{X_{\alpha_1 \alpha_2 \ldots \alpha_p}}{\qty(\frac{2n+p}{2})!}
    \tr \mathbbm{b}_{(2n)}^{\alpha_1 \alpha_2 \ldots \alpha_p}\,.
    \label{ck-gen-series}
\end{align}
In particular, the most important coefficients $c_1(x)$ and $c_2(x)$ can be expressed as the following series.
\begin{align}
c_1(x)
% &=
=
    \sum_{p=0}^\infty
    \frac{X_{\alpha_1 \alpha_2 \ldots \alpha_p}}{\qty(\frac{2+p}{2})!}
    \tr \mathbbm{b}_{(2)}^{\alpha_1 \alpha_2 \ldots \alpha_p}\,,
\qquad\qquad\qquad
c_2(x)
% &=
=
    \sum_{p=0}^\infty
        \frac{X_{\alpha_1 \alpha_2 \ldots \alpha_p}}{\qty(\frac{4+p}{2})!}
        \tr \mathbbm{b}_{(4)}^{\alpha_1 \alpha_2 \ldots \alpha_p}
\label{coeff-gen-series}\,,
\end{align}
where we have denoted the trace for matrix indices by $\tr\qty[\cdots]$.
To be more specific, for a matrix-valued tensor $\mathbbm{t}_{\mu_1 \cdots \mu_p}$, the trace $\tr\qty[\mathbbm{t}_{\mu_1 \cdots \mu_p}]$ is defined as
\begin{align}
\tr\qty[\mathbbm{t}_{\mu_1 \cdots \mu_p}] = \sum_{i=1}^{N}\qty(\mathbbm{t}_{\mu_1 \cdots \mu_p})_{ii}
\end{align}
where $\qty(\cdots)_{ij}$ means the $(i,j)$-component of the matrix.
These expressions are important to the calculation for 1-loop effective action in the two-dimensional case and the four-dimensional case.

\subsubsection{Calculation of the coefficients}
The explicit form of the coefficients $c_1(x)$ and $c_2(x)$ in \eqref{coeff-gen-series} are extremely complicated and cumbersome. However, we can reduce this computation by a certain special coordinate system, known as the Riemann normal coordinate. In this coordinate system, the covariant derivatives are reduced to the partial derivative and the metric tensor can be represented as a kind of Taylor expansion around a certain point \cite{Muller:1997zk, Nester2010NormalFF, Hartley:1995dg}.
Since their coefficients are covariant objects, we can recover the covariant form of the results even using the special coordinate.
We provide the important geometrical quantities and their derivatives on these coordinates in Appendix \ref{app:2_result-of-Seeley}. On the other hand, when we consider torsion and non-metricity, it is not possible to make them zero since they are tensors and do not generally vanish.
In the case that they are zero at a certain coordinate system, they are identically zero at any coordinate system due to the transformation law of tensors.
This means we should keep a finite value of the torsion and non-metricity tensors at the expanded point in the Taylor series.
After imposing the Riemann normal coordinate, many parts of the symbols' coefficients will vanish.
In particular, the series in \eqref{coeff-gen-series} are reduced to the following form:
\begin{align}
c_1(x)
&=  \Tr \mathbbm{b}_{(2)}^{\bullet}
    + \frac{1}{2!}
        X_{\alpha_1 \alpha_2}
        \Tr \mathbbm{b}_{(2)}^{\alpha_1 \alpha_2}
    + \frac{1}{3!}
        X_{\alpha_1 \alpha_2 \alpha_3 \alpha_4}
        \Tr \mathbbm{b}_{(2)}^{\alpha_1 \alpha_2 \alpha_3 \alpha_4}\,,
    \label{c1-gen-series-sp-coord} \\
c_2(x)
&=  \frac{1}{2!} \Tr \mathbbm{b}_{(4)}^{\bullet}
    + \frac{1}{3!} X_{\alpha_1 \alpha_2}
        \Tr \mathbbm{b}_{(4)}^{\alpha_1 \alpha_2}
    + \frac{1}{4!} X_{\alpha_1 \alpha_2 \alpha_3 \alpha_4}
        \Tr \mathbbm{b}_{(4)}^{\alpha_1 \alpha_2 \alpha_3 \alpha_4}\,,
    \label{c2-gen-series-sp-coord}
\end{align}
where a subscript $\bullet$ means that the quantity does not have indices. The form of each quantity is presented in the Appendix \ref{app-results-symbols-rnc}.
Finally, by substituting the equations in the Appendix \ref{app-results-symbols-rnc} into \eqref{c1-gen-series-sp-coord} and \eqref{c2-gen-series-sp-coord}, we find an explicit form of the coefficients $c_1(x)$ and $c_2(x)$ 
\begin{align}
c_1(x)
&=
    \Tr \qty( \mathbb{Z} + \frac{R}{6} \mathbbm{1}_N )\,,
    \label{eq:generic-result-c1}
    \\
c_2(x)
&=
    \frac{1}{6} \Tr \qty{\Box \qty(\mathbb{Z}+\frac{R}{5}\mathbbm{1}_N)}
    + \frac{1}{2} \Tr \qty{\qty(\mathbb{Z}+\frac{R}{6}\mathbbm{1}_N)^2}
\nonumber\\&\hspace{1cm}
    + \frac{1}{180} \Bigl( R_{\mu\nu\rho\sigma} R^{\mu\nu\rho\sigma} - R_{\mu\nu} R^{\mu\nu}\Bigr) \Tr \mathbbm{1}_N
    + \frac{1}{12} \Tr \qty(\mathbb{Y}_{\mu\nu}\mathbb{Y}^{\mu\nu})\,,
    \label{eq:generic-result-c2}
\end{align}
where we introduced the following tensors
\begin{align}
\begin{aligned}
\mathbb{Z} &=
    \mathbb{X} - \nabla_\mu \mathbb{S}^\mu - \mathbb{S}_\mu \mathbb{S}^\mu\,,
    \\
\mathbb{Y}_{\mu\nu} &=
    \mathbb{F}_{\mu\nu}
    + \nabla_\mu \mathbb{S}_\nu
    - \nabla_\nu \mathbb{S}_\mu
    + \mathbb{S}_\mu \mathbb{S}_\nu
    - \mathbb{S}_\nu \mathbb{S}_\mu
  \,,  \\
\mathbb{F}_{\mu\nu} \Psi &=
    \qty[ \tilde{\nabla}_\mu, \tilde{\nabla}_\nu ] \Psi\,.
\end{aligned}
\label{eq:tensors-in-coeffC}
\end{align}
Here, recall that tildes are denoting the general covariant derivative. Also, the vector $\mathbb{S}^\mu$ is defined by the coefficient of the covariant derivative for the Levi-Civita connection:
\begin{align}
    D = - \qty( g^{\mu\nu} \nabla_\mu \nabla_\nu + 2 \mathbb{S}^\mu \nabla_\mu + \mathbb{Z} )\,.
\end{align}
Let us emphasize here that the coefficients \eqref{eq:generic-result-c1} and \eqref{eq:generic-result-c2} are the final expressions derived from Seeley's algorithm, and they play a crucial role in the trace anomaly effect for the energy-momentum tensor. Thus, these expressions represent one of the most important results in this manuscript.%
\subsubsection{The commutator of covariant derivatives on fields}
Note that the tensor $\mathbb{F}_{\mu\nu}$ depends on the field we consider.
Using the spin connection $\tilde{\omega}_\mu$, this tensor is represented as
\begin{align}
\mathbb{F}_{\mu\nu} =
    \partial_\mu \tilde{\omega}_\nu - \partial_\nu \tilde{\omega}_\mu
    + \tilde{\omega}_\mu \tilde{\omega}_\nu - \tilde{\omega}_\nu \tilde{\omega}_\mu\,.
\end{align}
For instance, on a scalar field $\phi$ and a fermionic field $\psi$, the commutator of covariant derivatives acts as
\begin{align}
\qty[ \tilde{\nabla}_\mu, \tilde{\nabla}_\nu ] \phi
    &= 0
    \qquad \Longrightarrow \qquad \mathbb{F}_{\mu\nu} = 0\,,
    \label{CoV-commutator-scalar-EC} \\
\qty[ \tilde{\nabla}_\mu, \tilde{\nabla}_\nu ] \psi
    &= \frac{1}{4} \tilde{R}_{\rho\sigma\mu\nu} \sigma^{\rho\sigma} \psi
    \qquad \Longrightarrow   \qquad\mathbb{F}_{\mu\nu} = \frac{1}{4} \tilde{R}_{\rho\sigma\mu\nu} \sigma^{\rho\sigma}\,,
    \label{CoV-commutator-fermion-EC}
\end{align}
with $\sigma^{\mu\nu}=\frac{1}{2}\qty[\gamma^\mu,\gamma^\nu]$. When considering covariant derivatives $\nabla_\mu$ for the Levi-Civita connection or the spin connection $\omega_\mu$ including only the Riemannian part, a generic curvature tensor $\tilde{R}_{\mu\nu\rho\sigma}$ will be replaced by Riemannian one $R_{\mu\nu\rho\sigma}$ as follows. It is possible to make a post-Riemannian expansion and re-express the general covariant derivative in terms of Levi-Civita covariant derivative plus additional degrees of freedom coming from torsion and nonmetricity. Then, it would be also useful to recall that for the Riemannian sector, we have the following identity~\cite{Shapiro:2016pfm, Buchbinder:2021wzv}:
\begin{align}
    R_{\rho\sigma\mu\nu} \gamma^\mu \gamma^\nu \gamma^\rho \gamma^\sigma = - 2 R \label{Riemann-and-Gamma-matrices}
\end{align}
that would be useful later on.

\subsection{Coefficients in the scalar field theories on specific gravitational theories}
The advantage of Seeley's algorithm is that its scheme is systematic and independent of the space-time geometry.
Once we expand the differential operator $D$ in the local coordinate, we can straightforwardly compute the symbols and then the coefficients.
It implies that we can use the same symbols and equations by writing down the representations in the local coordinate even if considering another kind of spacetime geometry.
We briefly consider the application of the above results to a multiple field $\Psi$ under the Riemannian case \cite{Gilkey:1975iq}, thr Riemann-Cartan case \cite{Obukhov:1983mm} and then, its generalization to the generic MAG case.
For this purpose, we consider the following action:
\begin{align}
    S = \int d^4x \sqrt{-g}\, \frac{1}{2} \vec{\Psi}^\dagger D \vec{\Psi}\,,
\end{align}
where $D$ denotes the differential operator that would introduce differences in the Riemannian and non-Riemannian cases.

\subsubsection{Riemannian case}
Let us apply the above analysis to the Riemannian case \cite{Gilkey:1975iq}.
Here we consider the differential operator $D$ as follows:
\begin{align}
D =
    - \qty(
        g^{\mu\nu} \nabla_\mu \nabla_\nu \mathbbm{1}_{N}
        + \mathbb{A}^\mu \nabla_\mu
        + \mathbb{X}
    )\,,
\label{diff-op-D_GR}
\end{align}
where the covariant derivative is the same as the Riemannian covariant derivative $\nabla_\mu$ in this case.
By expanding the covariant derivatives, we obtain the representation of the coefficients in \eqref{expand_gen-cov-deriv_coord} as follows
\begin{align}
\mathbb{S}^\mu
&=
    \frac{1}{2} \mathbb{A}^\mu
  \,,  \\
\mathbb{H}^\mu
&=
    \mathbb{A}^\mu
    + \qty(
        2 g^{\mu\nu} \omega_\nu
        - g^{\alpha\beta} \Gamma^{\mu}_{\phantom{\mu}\beta\alpha}
    ) \mathbbm{1}_N\,,
    \label{coeff-H_GR}
    \\
\mathbb{E}
&=
    \mathbb{X}
    + g^{\mu\nu}
        \qty(
            \omega_\mu\omega_\nu
            + \partial_\mu\omega_\nu
            - \Gamma^{\alpha}_{\phantom{\alpha}\mu\nu} \omega_\alpha
        ) \mathbbm{1}_N
    + \mathbb{A}^\mu \omega_\mu\,.
    \label{coeff-E_GR}
\end{align}
Now we can derive the first two Seeley-DeWitt coefficients $c_1(x)$ and $c_2(x)$ by substituting \eqref{coeff-H_GR}, \eqref{coeff-E_GR}, into the expressions \eqref{c1-gen-series-sp-coord} and \eqref{c2-gen-series-sp-coord}. 

For instance, when we consider the case $D=-\qty(g^{\mu\nu}\nabla_\mu \nabla_\nu+\mathbb{X})$, we find the first coefficient $c_1(x)$ takes the following form
\begin{align}
c_1(x)
&=
    \tr \qty( \mathbb{X} + \frac{R}{6} \mathbbm{1}_N )\,.
\end{align}
This is the well-known result of the Seeley-DeWitt expansion \cite{Vassilevich:2003xt, Buchbinder:2021wzv}.

\subsubsection{Riemann-Cartan case}
We can easily extend the Seeley-DeWitt coefficients to the Riemann-Cartan case \cite{Obukhov:1983mm} based on Seeley's algorithm.
In Riemann-Cartan geometries, we consider the differential operator $D$:
\begin{align}
D =
    - \qty(
        g^{\mu\nu} \tilde{\nabla}_\mu \tilde{\nabla}_\nu \mathbbm{1}_{N}
        + \mathbb{A}^\mu \tilde{\nabla}_\mu
        + \mathbb{X}
    )\,,
\label{diff-op-D_EC}
\end{align}
where now the covariant derivative $\tilde{\nabla}$ is computed with the affine connections including the torsion degrees of freedom.
From the above equation, we can write down the coefficients $\mathbb{H}^\mu, \mathbb{S}_\mu$ and $\mathbb{E}$ in \eqref{expand_gen-cov-deriv} as follows: We decompose the general covariant derivative using a post-Riemannian expansion and then all the additional degrees of freedom coming from the non-Riemannian sector would be encoded in the vector $\mathbb{S}_\mu$. After doing that, the vector $\mathbb{S}_\mu$ behaves as
\begin{align}
\mathbb{S}^\mu
=
    \frac{1}{2} \qty(
        \mathbb{A}^\mu
        - g^{\rho\sigma}
            N^\mu_{\phantom{\mu}\rho\sigma}
            \mathbbm{1}_N
    )
=
    \frac{1}{2} \qty(\mathbb{A}^\mu + T^\mu \mathbbm{1}_N)\,,
\label{vectorS-EC}
\end{align}
and the other coefficients appearing in~\eqref{expand_gen-cov-deriv_coord} will be given by
\begin{align}
\mathbb{H}^\mu &=
    2 \mathbb{S}^\mu
    + \qty(
        2 g^{\mu\nu} \omega_\nu
        - g^{\alpha\beta} \Gamma^{\mu}_{\phantom{\mu}\alpha\beta}
    ) \mathbbm{1}_N\,, \\
\mathbb{E} &=
    \mathbb{X}
    + 2 \mathbb{S}^\mu \omega_\mu
    + g^{\mu\nu} \qty(
        \partial_\mu\omega_\nu
        + \omega_\mu\omega_\nu
        - \Gamma^\alpha_{\phantom{\alpha}\mu\nu} \omega_\alpha
    ) \mathbbm{1}_N\,.
\end{align}
Therefore, in this case, the vector $\mathbb{S}^\mu$ contains the trace part of the torsion tensor that is a vector. On the other hand, the spin connection $\tilde{\omega}^{A}{}_{B\mu}$ is related to the affine connection as follows:
\begin{align}
\tilde{\omega}^{A}{}_{B\mu}
=   e^{A}\,_{\lambda}\,e_{B}\,^{\rho}\,\tilde{\Gamma}^{\lambda}\,_{\rho \mu}+e^{A}\,_{\lambda}\,\partial_{\mu}\,e_{B}\,^{\lambda}\,.
\end{align}
Using these equation for \eqref{eq:generic-result-c1} and \eqref{eq:generic-result-c2}, the coefficients $c_1(x)$ and $c_2(x)$ will be found.
It is important to note that the covariant derivative must be considered whether it respects the Levi-Civita connection or an affine connection that includes torsion.

\subsubsection{Generic Metric-Affine Gravity case}
From the Riemann-Cartan case to the generic MAG case, the modification appears only in the covariant derivative since now the connection contains nonmetricity as well.
We recall that the advantage of Seeley's algorithm is that we can use the same scheme after expanding the differential operator $D$ in the local coordinate.
Therefore, we can easily extend the computations in the Riemann-Cartan case to the generic MAG case by redefining the vector $\mathbb{S}^\mu$ and the spin connection $\tilde{\omega}_\mu$ with torsion and nonmetricity.  Here, let us consider the same notations for the differential operator $D$ as \eqref{diff-op-D_EC} with the coefficient $\mathbb{H}^\mu$ and $\mathbb{E}$.
As seen in the previous case, we find the equations for coefficients $\mathbb{H}^\mu$ and $\mathbb{E}$ where the vector $\mathbb{S}^\mu$ is redefined by the following equation:
\begin{align}
\mathbb{S}^\mu
=
    \frac{1}{2} \qty(
        \mathbb{A}^\mu
        - g^{\rho\sigma}
            N^\mu_{\phantom{\mu}\rho\sigma}
            \mathbbm{1}_N
    )
=
    \frac{1}{2} \qty{
        \mathbb{A}^\mu
        + \qty(
            T^\mu
            + \frac{9}{4} \Lambda^\mu
            - W^\mu
        ) \mathbbm{1}_N
    }\,.
\label{vectorS-MAG}
\end{align}
Notice that the two traces of nonmetricity now enters in this equation.

\section{Applications: Trace anomaly for the energy-momentum tensor and hypermomentum} \label{sec:5}

In the following, we will consider applications for three different types of possible transformations related to the metric and the connection, with the aim of finding the trace anomalies and exploring possible extensions involving hypermomentum. 
\subsection{Scale transformations and invariance}
As one of the applications for our heat kernel method, we can straightforwardly consider the anomaly relevant to scaling transformations.
In MAG, we can define the following three types of scaling transformation \cite{Iosifidis:2018zwo,Sauro:2022hoh,Janssen:2018exh}.
\begin{itemize}
    \item {\bf Conformal transformation :} $g_{\mu\nu} \to e^{2\omega}g_{\mu\nu}$, $\tilde{\Gamma}^{\lambda}{}_{\mu\nu} \to \tilde{\Gamma}^{\lambda}{}_{\mu\nu}$
    \item {\bf Projective transformation :} $g_{\mu\nu} \to g_{\mu\nu}$, $\tilde{\Gamma}^{\lambda}{}_{\mu\nu} \to \tilde{\Gamma}^{\lambda}{}_{\mu\nu}+\delta^\lambda_\mu \xi_\nu$
    \item {\bf Frame rescaling :} $g_{\mu\nu} \to e^{2\omega}g_{\mu\nu}$, $\tilde{\Gamma}^{\lambda}{}_{\mu\nu} \to \tilde{\Gamma}^{\lambda}{}_{\mu\nu}+\delta^\lambda_\mu \partial_\nu\omega$
\end{itemize}
In the case of GR, the Levi-Civita connection is also transformed under the conformal transformation.
Since the metric tensor and the affine connection are independent of each other, we can define more transformations as expressed above.
Under an infinitesimal transformation for the metric and the affine connection, the action $S_{\rm m}$ for matter fields is transformed as:
\begin{align}
\delta S_{\rm m}
&=
  -  \frac{1}{2} \int d^4x \sqrt{-g}
        \qty{
            T_{\mu\nu} \delta g^{\mu\nu}
            + \Delta_\lambda^{\phantom{\lambda}\mu\nu}
                \delta \tilde{\Gamma}^\lambda_{\phantom{\lambda}\mu\nu}
        }\,,
\label{variation-of-classical-action}
\end{align}
where $ T_{\mu\nu}$ is the energy-momentum tensor and $\Delta_\lambda^{\phantom{\lambda}\mu\nu}$ is the hypermomentum tensor. In this equation, $\delta g_{\mu\nu}$ and $\delta\Gamma^{\lambda}_{\phantom{\lambda}\mu\nu}$ are infinitesimal variations of the metric and the connection.
Imposing the invariance under this transformation on the action $\delta S_{\rm m} = 0$, we find a relation on the stress-energy or/and hypermomentum tensor. For instance, when considering conformal invariant theories, this relation will be reduced to the equation for the traceless property.
On the other hand, at the quantum level, the variation for the effective action $\delta\Gamma$ would be
\begin{align}
\delta { {S_{\text{eff}}}}
=-\frac{1}{2} \int d^4x \sqrt{-g}
    \qty{
        \expval{T_{\mu\nu}} \delta g^{\mu\nu}
        + \expval{\Delta_\lambda^{\phantom{\lambda}\mu\nu}}
            \delta \tilde{\Gamma}^\lambda_{\phantom{\lambda}\mu\nu}
    }\,.
\label{variation-scaling-MAG}
\end{align}
As seen in the previous section, we can evaluate this variation using the heat kernel expansion.
Note that its result strongly depends on each theory and its action.
In the following subsections, we will consider and evaluate the variation of effective action with the scale-invariant theories based on the paper \cite{Iosifidis:2018zwo} \footnote{We note that the authors in the paper \cite{Iosifidis:2018zwo} use the opposite signature in the definition of the non-metricity tensor. So, if writing down non-metricity tensor by using covariant derivatives of the metric tensors, our $\mathcal{L}_{QT}$ has an opposite signature with respect to the paper \cite{Iosifidis:2018zwo}.}.
We consider a single scalar field $\phi$ and the following action $S$
\begin{align}
S&=
    \int\sqrt{-g}{d^4 x}
    \lambda \left[
        \mathcal{L}_Q
        + \mathcal{L}_T
        + \mathcal{L}_{QT}
    \right] \phi^2
    + S_\phi\,,
    \\
S_\phi&=
    \int\sqrt{-g}{d^4x}
        \frac{1}{2} \phi \mathcal{O} \phi\,,
\qquad \textrm{with}\qquad    
        \mathcal{O} = - ( g^{\mu\nu} \nabla_\mu \nabla_\nu + 2 S^\mu \nabla_\mu + Z )
    \,,
\end{align}
where $\mathcal{O}$ is a second-order differential operator and each $\mathcal{L}_i$ are defined as the linear combination of quadratic terms for torsion and nonmetricity:
\begin{align}
\mathcal{L}_Q
&=
    p_1 Q_{\alpha\mu\nu} Q^{\alpha\mu\nu}
    + p_2 Q_{\alpha\mu\nu} Q^{\mu\nu\alpha}
    + p_3 Q_\mu Q^\mu
    + p_4 \hat{Q}_\mu \hat{Q}^\mu
    + p_5 Q_\mu \hat{Q}^\mu\,,
    \\
\mathcal{L}_T
&=
    q_1 T_{\alpha\mu\nu} T^{\alpha\mu\nu}
    + q_2 T_{\alpha\mu\nu} T^{\mu\nu\alpha}
    + q_3 T_\mu T^\mu\,,
    \\
\mathcal{L}_{QT}
&=
    r_1 Q_{\mu\nu\lambda} T^{\lambda\mu\nu}
    + r_2 Q_\mu T^\mu
    + r_3 \hat{Q}_\mu T^\mu\,.
\end{align}
Here, we defined the vectors  $Q_\mu=Q_{\mu\alpha\lambda} g^{\alpha\lambda}$, and $\hat{Q}_\mu=Q_{\lambda\alpha\mu} g^{\lambda\alpha}$ to match the convention of \cite{Iosifidis:2018zwo}. It is easy to see that those vectors are related to ours as
\begin{eqnarray}
    Q_\mu=4W_\mu \,,\quad \hat{Q}=\frac{9}{4}\Lambda_\mu+W_\mu\,.
\end{eqnarray}
Then, there are $5 + 3 + 3 = 11$ coefficients ($p_i$, $q_i$ and $r_i$) that parametrizes the action.
In what follows, after requiring an invariance of this action under a particular transformation, these coefficients will be constrained by certain conditions.
Note that we use non-bold characters for the coefficients $A^\mu$ and $X$ since a scalar field $\phi$ is singlet in this case.

\subsubsection{Conformal transformation}
In the conformal transformation, the variation of the metric and connection are given by
\begin{align}
    \delta g_{\mu\nu} = 2 \omega g_{\mu\nu}
    \qquad
    \delta \tilde{\Gamma}^\lambda_{\phantom{\lambda}\mu\nu} = 0\,,
\label{eq:infsimal-conformal}
\end{align}
and a scalar field is transformed as
\begin{align}
    \phi \to e^{-\omega} \phi\,.
\end{align}
Imposing the invariance under conformal transformation, the parameters are constrained by the following equations (in a four-dimensional spacetime):
\begin{align}
\begin{aligned}
0=& 2 p_1 + 8 p_3 + p_5\,, \\
0=& 2 p_2 + 2 p_4 + 4 p_5 \,,\\
0=& 4 p_1 + p_2 + 16 p_3 + p_4 + 4 p_5 \,,\\
0=& r_1 + 4 r_2 + r_3\,.
\end{aligned}
\label{eq:parameter-for-conf-inv-action}
\end{align}
Note that, when imposing the conditions \eqref{eq:parameter-for-conf-inv-action}, the sum of $\mathcal{L}_i$ are invariant under projective transformation:
\begin{align}
\delta\qty( \mathcal{L}_Q + \mathcal{L}_T + \mathcal{L}_{QT} ) = 0\,.
\end{align}
Also, we can choose the action for the matter field with the operator $\mathcal{O}$ as follows
\begin{align}
S_\phi
&=
- \frac{1}{2}\int\sqrt{-g}{d^4 x}~
    \phi
    \left[
        g^{\mu\nu} \nabla_\mu \nabla_\nu
        + \frac{1}{8} \qty(\nabla_\lambda Q^\lambda)
        - \frac{1}{64} Q_\lambda Q^\lambda
    \right]
    \phi
\end{align}
that corresponds to having
\begin{align}
S^\mu = 0
\qquad
Z =
    \qty{
        \frac{1}{8} \qty(\nabla_\lambda Q^\lambda)
        - \frac{1}{64} Q_\lambda Q^\lambda
        - 2 \qty( \mathcal{L}_Q + \mathcal{L}_T + \mathcal{L}_{QT} )
    } \,\mathbbm{1}_N
\label{eq:coeff-conformal-inv}
\end{align}
in Eq.~\eqref{gen-cov-deriv}.
We can see that the theory with the action $S=S_g+S_\phi$ is a conformally invariant theory with $\phi \to \phi$.
Classically, the invariance under conformal transformation leads to the traceless property of energy-momentum tensor:
\begin{align}
    {T_\mu}^\mu = 0\,,
\end{align}
This is the same as the traceless property in the conformal invariant theory within GR (or Riemannian geometry).
At the quantum level, applying \eqref{eq:infsimal-conformal} to \eqref{variation-scaling-MAG} we find the variation of the effective action as follows:
\begin{align}
\delta{ {S_{\text{eff}}}}
=\int d^4x \sqrt{-g} \expval{T_{\mu}^{\phantom{\mu}\mu}} \omega(x)\,.
\end{align}
This is the same result as the Weyl anomaly in the Riemannian (GR) case.
Thus, we can find the anomaly on the trace of the energy-momentum tensor by using the coefficient $c_2(x)$:
\begin{align}
    \expval{T_{\mu}^{\phantom{\mu}\mu}} = \frac{c_2(x)}{(4\pi)^2}\,,
\end{align}
where the coffiecient $c_2(x)$ is determined by the equation \eqref{eq:generic-result-c2} with the coefficients \eqref{eq:tensors-in-coeffC}, \eqref{vectorS-MAG}, and \eqref{eq:coeff-conformal-inv}, namely
\begin{align}
c_2(x)
&=
    \frac{1}{6} \qty{\Box \qty(
        \frac{1}{8} \qty(\nabla_\lambda Q^\lambda)
        - \frac{1}{64} Q_\lambda Q^\lambda
        - 2 \qty( \mathcal{L}_Q + \mathcal{L}_T + \mathcal{L}_{QT} )
        + \frac{R}{5}
    )}
\nonumber\\&\hspace{1cm}
    + \frac{1}{2} \qty{\qty(
        \frac{1}{8} \qty(\nabla_\lambda Q^\lambda)
        - \frac{1}{64} Q_\lambda Q^\lambda
        - 2 \qty( \mathcal{L}_Q + \mathcal{L}_T + \mathcal{L}_{QT} )
        + \frac{R}{6}
    )^2}
\nonumber\\&\hspace{2cm}
    + \frac{1}{180} \Bigl( R_{\mu\nu\rho\sigma} R^{\mu\nu\rho\sigma} - R_{\mu\nu} R^{\mu\nu}\Bigr)\,.\label{anomaly1}
\end{align}
Then,  in a conformally invariant theory for MAG, the so-called trace anomaly is modified by the presence of the traces of torsion and nonmetricity.
The difference with the GR case is that the coefficient $c_2(x)$ contains the torsion and non-metricity tensors. 
So, this trace anomaly on the stress-energy tensor gets the non-Riemannian contributions as the quantum correction.

\subsubsection{Projective transformation}
In the projective transformation, the variation of the metric and connection are given by
\begin{align}
    \delta g_{\mu\nu} = 0
    \qquad
    \delta \tilde{\Gamma}^\lambda_{\phantom{\lambda}\mu\nu} = \delta^\lambda_\mu \xi_\nu\,,
\label{eq:infsimal-proj}
\end{align}
and a scalar field is not transformed, i.e.,
\begin{align}
    \phi \to \phi\,.
\end{align}
This case does not exist in the Riemnaninan case since it is not possible to transforms only the connection. 
Imposing the invariance of the action $S=S_g+S_\phi$ under this transformation \eqref{eq:infsimal-proj}, we find the following constraint 
\begin{align}
\begin{aligned}
0&=
    4 \qty(  2 p_1 + p_5 + 8 p_3 )
    + \qty( r_1 + 3 r_2 )
 \,,   \\
0&=
    4 \qty( 2 p_2 + 2 p_4 + 4 p_5 )
    - \qty( r_1 - 3 r_3 )\,,
    \\
0&=
    2 q_1 - q_2 + 3 q_3
    + 2 \qty( r_1 + 4 r_2 + r_3 )\,,
    \\
0&=
    16 \qty( 4 p_1 + p_2 + 16 p_3 + p_4 + 4 p_5 )
    + 3
        \qty{
            2 q_1 - q_2 + 3 q_3
            + 4 \qty( r_1 + 4 r_2 + r_3 )
        }\,.
\end{aligned}
\label{eq:parameter-for-proj-inv-action}
\end{align}
Further, let us choose the scalar field action $S_\phi$ as follows:
\begin{align}
S_\phi
    = \frac{1}{2} \int\sqrt{-g} d^4x \qty( -\phi g^{\mu\nu} \nabla_\mu \nabla_\nu \phi )\,.
\end{align}
This case corresponds to $S^\mu = 0$ and $Z = -2\qty(\mathcal{L}_Q+\mathcal{L}_T+\mathcal{L}_{QT})$ in Eq.~\eqref{gen-cov-deriv}.
The projective invariance of the theory ($\delta S=0$) leads to the traceless property on the hypermomentum tensor from the equation \eqref{variation-of-classical-action}:
\begin{align}
\delta S
&=
   - \frac{1}{2} \int d^4x \sqrt{-g}
        \qty{
            \Delta_\lambda^{\phantom{\lambda}\mu\nu}
            \qty(\delta^\lambda_\mu \xi_\nu)
        }=
   - \frac{1}{2} \int d^4x \sqrt{-g}
        \qty{
            \Delta_\lambda^{\phantom{\lambda}\lambda\nu} \xi_\nu
        }
    = 0\quad 
    \Longrightarrow
    \Delta_\lambda^{\phantom{\lambda}\lambda\mu} = 0\,,
\end{align}
where we used the fact that the vector \(\xi_\mu\) is arbitrary in the last step. This implies that the dilation component of the hypermomentum tensor must vanish under projective invariance, i.e., \(^{(\text{d})}\Delta_\mu = 0\) (see Eq.~\eqref{hypermomentum} for the decomposition of the hypermomentum tensor). At the quantum level, applying \eqref{eq:infsimal-proj} to \eqref{variation-scaling-MAG}, the variation of the effective action becomes
\begin{align}
\delta{{S_{\text{eff}}}}
=\int d^4x \sqrt{-g}
    \expval{\Delta_\lambda^{\phantom{\lambda}\lambda\nu}}
    \xi_\nu
=\int d^4x \sqrt{-g}
    \expval{{}^{(\text{d})}\Delta^\mu} \xi_\mu\,.
\label{variation-eff-action-with-proj-inv}
\end{align}
Thus, the variation of the effective action in the case of projective invariance gives rise to a quantity that depends solely on the hypermomentum tensor. This is a generic result for a projective invariant theory.
Let us consider the quantum effect for the traceless property in this projective invariant theory.
From the above discussion, a theory constructed from the action $S=S_g+S_\phi$ is invariant under projective transformations with $\phi \to \phi$.
Because of the invariance of the field $\phi$, the metric $\delta g_{\mu\nu}=0$, and the action $\delta S=0$, we find the variation $\delta\mathcal{O}$ of the operator $\mathcal{O}$ from the variation of the action $\delta S$ as follows
\begin{align}
\delta S
&=
    \delta\qty[
        \int\sqrt{-g} ~ \frac{1}{2} \phi \mathcal{O} \phi
    ]\,,
    \\
&=
    \int\sqrt{-g} ~ \frac{1}{2} \phi \qty(\mathcal{O}+\delta\mathcal{O}) \phi
    - \int\sqrt{-g} ~ \frac{1}{2} \phi \mathcal{O} \phi
 \,,   \\
&=
    \int\sqrt{-g} ~ \frac{1}{2} \phi \qty(\delta\mathcal{O}) \phi
= 0\,.
\end{align}
To find the effective action, let us go back to Eq.~\eqref{variation-via-heat-kernel} using the proper time representation. By substituting $\delta\mathcal{O}=0$, we find that the variation of the effective action vanishes
\begin{align}
\delta{ {S_{\text{eff}}}}[g_{\mu\nu}]
=   \frac{1}{2} \int^\infty_{1/\Lambda^2} ds \Tr \qty(e^{-s\mathcal{O}}\delta\mathcal{O})
=   0\,.
\end{align}

Compared with \eqref{variation-eff-action-with-proj-inv}, this equation implies that the variation of the effective action will also vanish, $\delta\Gamma = 0$.
Thus the expectational value of the hypermomentum is zero:
\begin{align}
\expval{{}^{(\text{d})}\Delta^\mu} = 0\,.
\end{align}
This expression follows from the invariance of the metric $\delta g_{\mu\nu}=0$ under the transformation. This implies that at one-loop level, we need to consider a transformation for the metric to obtain quantum anomalies from the variation of the effective action.%

\subsubsection{Other projective-invariant theories}
In the previously studied projective invariant theory, no anomaly was found for theories of the type presented in \cite{Iosifidis:2018zwo}. Let us now consider another theory that is projective invariant and see if any anomaly appears or not.
Inspired by \cite{Janssen:2018exh}, let us introduce the following action of a fermion field $\psi$
\begin{align}
S
=
    \int{d^4x}\sqrt{-g}~\qty{
        \frac{i}{2}
            \bar{\psi} ~\gamma^\mu \qty( \tilde{\nabla}_\mu + \frac{i}{2} W_\mu ) \psi
        + \text{h.c.}
    }\,,
\end{align}
where $W_\mu$ is the Weyl vector.
For this action, we define the scaling transformation as follows
\begin{align}
    \psi \longrightarrow e^{i\Omega} \psi\,,
    \qquad
    \tilde{\Gamma}^\lambda{}_{\mu\nu} \longrightarrow \tilde{\Gamma}^\lambda{}_{\mu\nu} + {\delta^\lambda}_\mu \partial_\nu \Omega\,,
    \qquad
    W_\mu \longrightarrow W_\mu - 2 \partial_\mu \Omega\,,
    \qquad
    g_{\mu\nu} \longrightarrow g_{\mu\nu} \,,
\label{eq:another-proj-inv-trans}
\end{align}
which is a subclass of the projective transformation. By construction, the above action is  invariant under this transformation law \eqref{eq:another-proj-inv-trans}.
In contrast to the previous section, the scaling transformation also affects the fermionic field. The projective invariance $\delta S_{\rm m}$ under this transformation \eqref{eq:another-proj-inv-trans} leads the traceless property on the corresponding hypermomentum tensor $\Delta_{\lambda\mu\nu}$:
\begin{align}
\delta S
&=
    \int d^4x \sqrt{-g}~
        \frac{\delta S_{\rm m}}{\delta \tilde{\Gamma}^\lambda{}_{\mu\nu}}
        \delta \tilde{\Gamma}^\lambda{}_{\mu\nu}
    \\
&=
    \int d^4x \sqrt{-g}~
        {\Delta_\lambda}^{\mu\nu}
        \qty( {\delta^\lambda}_\mu \partial_\nu \Omega )
    \\
&=
    \int d^4x \sqrt{-g}~
      -  \qty{
            \frac{1}{\sqrt{-g}}
            \partial_\mu \qty( \sqrt{-g} {\Delta_\lambda}^{\lambda\mu} )
        }
        \Omega
=
    0
    \\
&\Longrightarrow
    \frac{1}{\sqrt{-g}} \partial_\mu \qty( \sqrt{-g} ~ {}^{(\text{d})}\Delta^\mu ) =
    \nabla_\mu {}^{(\text{d})}\Delta^\mu = 0\,.
\end{align}
The above equation implies that, in this case, the dilation part ${}^{(\text{d})}\Delta^\mu$ of the hypermomentum tensor is covariantly conserved (with respect to the Levi-Civita connection).
Let us now find out the effective action based on this transformation law \eqref{eq:another-proj-inv-trans}.
From the transformation law on a fermionic field $\psi$, the operator $\mathcal{O}$ is transformed to $\mathcal{O}+\delta\mathcal{O}$ as
\begin{align}
    \delta \mathcal{O} = i \qty( \Omega \mathcal{O} - \mathcal{O} \Omega )
\end{align}
up to the infinitesimal transformation. By applying this equation to the variation of the effective action, we find that the action at one-loop effect becomes
\begin{align}
\delta{ {S_{\text{eff}}}}
&=
    -\frac{1}{2}
    \int^\infty_{1/\Lambda^2} ds~
        \Tr\qty[
            e^{-s\mathcal{O}} \qty{i \qty( \Omega \mathcal{O} - \mathcal{O} \Omega )}
        ]=
    -\frac{i}{2}
    \int^\infty_{1/\Lambda^2} ds~
        \Tr\qty[
            e^{-s\mathcal{O}} \Omega \mathcal{O}
            - e^{-s\mathcal{O}} \mathcal{O} \Omega
        ]=
    0\,,
\end{align}
where we have used the cyclic property of the (functional) trace in the last line. Similarly to what occurred in the previous section, there is no trace anomaly at the one-loop level for this theory. Note that in this case, even though the field is transformed, there is still no effect concerning anomalies.

\subsubsection{Frame rescaling}
In the frame rescaling, the variations for the metric and connection sectors become
\begin{align}
    \delta g_{\mu\nu} = - 2\omega g_{\mu\nu}
    \qquad
    \delta {\tilde{\Gamma}^\lambda}_{\mu\nu} = \delta^\lambda_\mu \partial_\nu\omega\,,
\label{eq:infsimal-frame-rescaling}
\end{align}
and a scalar field is transformed as
\begin{align}
    \phi \to e^{-\omega} \phi\,.
\end{align}
We require the following conditions to obtain an invariant theory with an action $S=S_g+S_\phi$ under the transformation \eqref{eq:infsimal-frame-rescaling}:
\begin{align}
0&= 2 q_1 - q_2 + 3 q_3\,, \quad 0= r_1 + 3 r_2\,,\quad 0= r_1 - 3 r_3\,.
\label{eq:parameter-for-rescaling-inv-action}
\end{align}
Also, as we have seen in the previous cases, the invariance under the frame rescaling determines the action $S_\phi$.
If we impose that a scalar field $\phi$ is transformed as $\phi \to e^{-\omega} \phi$, we find 
\begin{align}
S=
-\frac{1}{2}\int\sqrt{-g}{d^4 x}
    \phi
    \qty{
        g^{\mu\nu} \nabla_\mu \nabla_\nu
        - \frac{2}{3} \nabla_\lambda T^\lambda
        + \frac{4}{9} T^\mu T_\mu
    }
    \phi\,.
\end{align}
From this action, the coefficients $S^\mu$ and $Z$ yield
\begin{align}
S^\mu
=
    0
\qquad
Z
=
    - \frac{2}{3} \nabla_\lambda T^\lambda
    + \frac{4}{9} T^\mu T_\mu
    - 2 \qty(\mathcal{L}_Q + \mathcal{L}_T + \mathcal{L}_{QT})\,,
\end{align}
and then the coefficient $c_2$ in this case as yields
\begin{align}
c_2(x)
&=
    \frac{1}{6} \qty{\Box \qty(
        - \frac{2}{3} \nabla_\lambda T^\lambda
    + \frac{4}{9} T^\mu T_\mu
    - 2 \qty(\mathcal{L}_Q + \mathcal{L}_T + \mathcal{L}_{QT})
        + \frac{R}{5}
    )}
\nonumber\\&\hspace{1cm}
    + \frac{1}{2} \qty{\qty(
        - \frac{2}{3} \nabla_\lambda T^\lambda
    + \frac{4}{9} T^\mu T_\mu
    - 2 \qty(\mathcal{L}_Q + \mathcal{L}_T + \mathcal{L}_{QT})
        + \frac{R}{6}
    )^2}
\nonumber\\&\hspace{2cm}
    + \frac{1}{180} \Bigl( R_{\mu\nu\rho\sigma} R^{\mu\nu\rho\sigma} - R_{\mu\nu} R^{\mu\nu}\Bigr)\,.
\end{align}
Under the frame rescaling, the invariance of the theory ($\delta S=0$) leads to the traceless property on the hypermomentum tensor from the equation \eqref{variation-of-classical-action}, namely
\begin{align}
\delta S
&=
  -  \frac{1}{2} \int d^4x \sqrt{-g}
        \qty{
            T_{\mu\nu} \qty(- 2\omega g_{\mu\nu})
            + \Delta_\lambda^{\phantom{\lambda}\mu\nu}
                \qty(\delta^\lambda_\mu \partial_\nu\omega)
        }
    \\
&=
    -\frac{1}{2} \int d^4x \sqrt{-g}
        \qty{
            - 2 T^\mu_{\phantom{\mu}\mu} \omega
            + \Delta_\lambda^{\phantom{\lambda}\lambda\mu}
                \qty(\partial_\mu\omega)
        }
    \\
&=
    \int d^4x \sqrt{-g}
        \qty{
            T^\mu_{\phantom{\mu}\mu}
            + \frac{1}{2}
                \frac{1}{\sqrt{-g}}
                \partial_\mu\qty(
                    \sqrt{-g} {}^{(\text{d})}\Delta^\mu
                )
        } \omega
    \\
&\Longrightarrow
    T^\mu_{\phantom{\mu}\mu}
    + \frac{1}{2} \nabla_\mu {}^{(\text{d})}\Delta^\mu
    = 0\,.
\end{align}
Similarly, the variation of the effective action becomes
\begin{align}
\delta {S_{\text{eff}}}
&=
    \int d^4x \sqrt{-g}
        \qty{
            \expval{T^\mu_{\phantom{\mu}\mu}}
            + \frac{1}{2} \nabla_\mu \expval{ {}^{(\text{d})}\Delta^\mu }
        } \omega\,,
\end{align}
where we have integrated by part.
From the expansion of the heat kernel, we can obtain the variation of the effective action in the same way as the conformal transformation case. By doing that, we find
\begin{align}
\expval{T^\mu_{\phantom{\mu}\mu}} + \frac{1}{2} \nabla_\mu \expval{ {}^{(\text{d})}\Delta^\mu }
= \frac{c_2(x)}{(4\pi)^2} \,.
\end{align}
Thus, in this case, we not only find the standard trace anomaly for the energy-momentum tensor, but we also discover that one of the traces of the hypermomentum (the one related to dilations) appears in the above equation. Then, we found a trace anomaly related to the dilation part of the hypermomentum. This is a new effect that, to our knowledge, has not been found elsewhere.

\subsection{Theory for spin $1/2$}
In this section, we consider couplings betwern fermionic field $\psi$ with gravity in the MAG framework. Let us consider the kinetic term in the fermionic sector as~\cite{Rigouzzo:2023sbb}
\begin{align}
\mathcal{L}_{\text{kin}}=
\frac{i}{2} \bar{\psi} \gamma^\mu \qty(1-i\alpha-i\beta\gamma_5) \tilde{\nabla}_\mu \psi + \text{h.c.}
% \qquad
% \tilde{\mathcal{D}}_\mu
%     = \tilde{\nabla}_\mu - i \mathcal{A}_\mu + i \mathcal{B}_\mu \gamma^5
\label{def:non-minimal-coupling}\,,
\end{align}
with constants $\alpha$ and $\beta$ representing nonminimal couplings between torsion/nonmetricity and a spin $1/2$, respectively. Let us emphasise again that the above coupling is not the most general that one can encounter in MAG since the symmetric part of the connection does not couple to a spin $1/2$ field.  The decomposition of this kinetic Lagrangian into the Riemannian part and torsion/non-metricity part is
\begin{align}
\mathcal{L}_{\text{kin}}
&=  \qty(
        \frac{i}{2} \bar{\psi} \gamma^\mu \nabla_\mu \psi + \text{h.c.}
    )
    + \frac{\alpha}{4} \qty(2T_\mu+3W_\mu-\frac{9}{4}\Lambda_\mu) \bar{\psi} \gamma^\mu \psi
\nonumber\\&\qquad\qquad
    + \frac{\beta}{4} \qty(2T_\mu+3W_\mu-\frac{9}{4}\Lambda_\mu) \bar{\psi} \gamma_5 \gamma^\mu \psi
    - \frac{1}{8} S_\mu
     \bar{\psi} \gamma_5 \gamma^\mu \psi\,.
\end{align}
In the above equation, the case with $\alpha=\beta=0$ corresponds to the minimal fermion coupling in MAG.
Let us try to apply our heat kernel method for this theory and to derive the coefficient $c_2(x)$.
For simplicity in the computations, we redefine this kinetic term as follows.
\begin{align}
% \mathcal{L}_{\text{kin}}
% &=   i \bar{\psi} \gamma^\mu \qty(\tilde{\mathcal{D}}_\mu - iP_\mu + iR_\mu \gamma_5) \psi
%    \,, \\
\mathcal{L}_{\text{kin}}
&=   i \bar{\psi} \gamma^\mu \qty(
        % \tilde{\mathcal{D}}_\mu
        \nabla_\mu
        - i \mathcal{A}_\mu
        + i \mathcal{B}_\mu \gamma_5
    ) \psi
   \,, \\
\mathcal{A}_\mu
&   = \alpha V_\mu
   \,, \label{EqP}\\
\mathcal{B}_\mu 
&=  - \frac{1}{8} S_\mu + \beta V_\mu\,,\label{EqR}
    \\
V_\mu 
&=
    \frac{1}{4}
        \qty(
            2 T_\mu
            - 3 W_\mu
            + \frac{9}{4} \Lambda_\mu
        ) \,.\label{EqV}
\end{align}
In the Riemann-Cartan case, only torsion appears and couples with spin $1/2$ fields. The trace anomaly in this case was already reported in \cite{DeBerredo-Peixoto:2001wkv, Camargo:2022gcw, Shapiro:2001rz, Obukhov:1983mm}.

To apply our heat kernel method, we have to make the corresponding second-order differential operator $\mathcal{O}$ on the fermion field $\psi$.
In the standard way, we derive such an operator by taking the square of the kinetic operator, yielding
\begin{align}
\mathcal{O} \psi
&=
    - \left[
        g^{\mu\nu} \nabla_\mu \nabla_\nu
        + \qty(
            - 2 i \mathcal{A}^\mu
            + 2 i \gamma^{[\mu} \gamma^{\nu]} \gamma^5 \mathcal{B}_\nu
        ) \nabla_\mu
        \phantom{\frac{1}{8}}
\right.\nonumber\\&\qquad\qquad\left.
        + \left\{
            - i \nabla_\mu \mathcal{A}^\mu
            + i \qty(\nabla_\mu \mathcal{B}^\mu) \gamma^5
            - \mathcal{A}_\mu \mathcal{A}^\mu
            + \mathcal{B}_\mu \mathcal{B}^\mu
            - i \qty( \nabla_{\mu} \mathcal{A}_{\nu} ) \gamma^{[\mu} \gamma^{\nu]}
\right.\right.\nonumber\\&\qquad\qquad\qquad\qquad\qquad\qquad\left.\left.
            + i \qty( \nabla_{\mu} \mathcal{B}_{\nu}) \gamma^{[\mu} \gamma^{\nu]} \gamma^5
            + 2 \mathcal{A}_\mu \mathcal{B}_\nu \gamma^{[\mu} \gamma^{\nu]} \gamma^5
            - \frac{1}{4} R \mathbbm{1}_N
        \right\}
    \right] \psi\,.
\label{eq:second-order-generic-kinetic-operator}
\end{align}
If we compare the above expression with the definitions of the tensor $S^\mu$ and $X$, we find
\begin{align}
\mathcal{O} =
    -\qty(
        g^{\mu\nu} \nabla_\mu \nabla_\nu
        + 2 S^\mu \nabla_\mu
        + X
    )\,,
\end{align}
where
\begin{align}
S^\mu
&=  
    - i \mathcal{A}^\mu
    + i \gamma^{[\mu} \gamma^{\nu]} \gamma^5 \mathcal{B}_\nu\,,
    \\
Z
&=
    - i \nabla_\mu \mathcal{A}^\mu
    + i \qty(\nabla_\mu \mathcal{B}^\mu) \gamma^5
    - \mathcal{A}_\mu \mathcal{A}^\mu
    + \mathcal{B}_\mu \mathcal{B}^\mu
    - i \qty( \nabla_{\mu} \mathcal{A}_{\nu} ) \gamma^{[\mu} \gamma^{\nu]}
\nonumber\\&\qquad\qquad
    + i \qty( \nabla_{\mu} \mathcal{B}_{\nu}) \gamma^{[\mu} \gamma^{\nu]} \gamma^5
    + 2 \mathcal{A}_\mu \mathcal{B}_\nu \gamma^{[\mu} \gamma^{\nu]} \gamma^5
    - \frac{1}{4} R \mathbbm{1}_N\,.
\end{align}
From these equations, we can find the coefficient $c_2(x)$, leading to the one-loop correction in this theory. This term becomes
\begin{align}
c_2(x)
&=
    - \frac{1}{30} \Box R
    - \frac{1}{45} R_{\mu\nu} R^{\mu\nu}
    - \frac{7}{360} R_{\mu\nu\rho\sigma} R^{\mu\nu\rho\sigma}
    + \frac{1}{72} R^2
    - \frac{4}{3} \Box\qty(\mathcal{B}_\lambda \mathcal{B}^\lambda)
    + \frac{4}{3} \nabla_\mu\qty(
        \mathcal{B}^\nu \nabla_\nu \mathcal{B}^\mu
        - \mathcal{B}^\mu \nabla_\nu \mathcal{B}^\nu
    )
\nonumber\\&\qquad
    + \frac{2}{3}
        \qty( \nabla_\mu \mathcal{B}_\nu - \nabla_\nu \mathcal{B}_\mu )
        \qty( \nabla^\mu \mathcal{B}^\nu - \nabla^\nu \mathcal{B}^\mu )
    + \frac{2}{3}
        \qty( \nabla_\mu \mathcal{A}_\nu - \nabla_\nu \mathcal{A}_\mu )
        \qty( \nabla^\mu \mathcal{A}^\nu - \nabla^\nu \mathcal{A}^\mu )\,.
\end{align}
This coefficient $c_2(x)$ gives a one-loop divergence and an expectation value of the trace of the stress-energy tensor for this theory in a four-dimensional spacetime.
For instance, when imposing the conformal invariance, we find
\begin{align}
    \expval{{T_\mu}^\mu} = \frac{c_2(x)}{(4\pi)^2}\,.
\end{align}
We note that the coefficient $c_n(x)$ is needed if considering a one-loop contribution in the $2n$-dimensional spacetime.  Recall that \(\mathcal{A}_\mu\) and $\mathcal{B}_\mu$
are given by~\eqref{EqP} and ~\eqref{EqR} along with~\eqref{EqV}, and therefore depend on both torsion and nonmetricity. Thus, for the spin $1/2$ case, both torsion and nonmetricity directly affect the trace anomaly of the energy-momentum tensor.%

\section{Conclusions}\label{sec:conclusion}

In this paper, we have investigated the trace anomalies for scaling-invariant theories within the framework of Metric-Affine Gravity (MAG).
The first part of our work focuses on the generalization of the asymptotic expansion of the heat kernel, specifically the so-called HMDS coefficients.
By employing Seeley's algorithm, as used by Obukhov in his work, we systematically extended these coefficients.
As a result, the non-metricity and torsion tensors appear in the HMDS coefficients through the tensor \(\mathbb{S}^\mu\) and the spin connection.
In the second part of our paper, we explored some concrete applications of the generalized coefficients to scale-invariant scalar field theories and spin $1/2$ theory.
In particular, we demonstrated that among the three types of scaling transformations for a scalar field within MAG, only the projective-invariant case does not produce a trace anomaly at the one-loop level. While this absence of an anomaly for the projective transformation may generally hold, a complete (dis)proof is left for future work.
Additionally, the theory invariant under frame rescaling exhibits an anomaly in the relationship between the stress-energy tensor and the hypermomentum tensor. We also found that in certain cases, both torsion and nonmetricity modify the \(c_2\) coefficient appearing in the trace anomaly, potentially inducing new effects related to these quantities, similar to those studied in the Riemannian sector~\cite{Christensen:1977jc,Camargo:2022gcw,Mottola:2022tcn,Komargodski:2011xv}.

As one of the future directions, we should consider the general properties of the trace anomaly.
Now, in particular, we have shown several examples for evaluating the trace anomaly with scale-invariant theories.
Curiously, we have no anomaly in the projective invariant cases.
We indeed consider only a few examples of such invariant theories, so this feature is not guaranteed in general.
However, we consider it maybe robust even in other projective invariant theories and more generic cases.
Since projective invariance exists in Riemannian geometry (as in GR), but within a theory where an affine connection is introduced, we conjecture that scale invariance under projective transformation represents a different form of invariance compared to invariance under conformal transformation and frame rescaling. We need to confirm these points and discuss proof if it is true.
Another potential direction is the application of our results to theories involving non-metricity.
As an extension of the Riemannian case, Riemann-Cartan geometries with torsion have been extensively studied by several authors.
However, in the presence of non-metricity, many features remain unclear compared to the torsional case.
We believe that our results can provide insights into the quantum aspects of non-metricity, at least at the one-loop level.

\section*{Acknowledgements}
The authors express their gratitude to Yuri N. Obukhov for his extensive feedback and valuable comments on his previous work \cite{Obukhov:1983mm}.
S.B. is supported by “Agencia Nacional de Investigación y Desarrollo” (ANID), Grant “Becas Chile postdoctorado
al extranjero” No. 74220006. Y.~M. acknowledges the financial support from the Advanced Research Center for Quantum Physics and Nanoscience, Tokyo Institute of Technology, at the initial stage of this work and JSPS KAKENHI Grant Number 22KJ1332.
M.Y. is supported by IBS under the project code IBS-R018-D3 and JSPS KAKENHI Grant Numbers JP21H01080.

\appendix

\section{The result of Seeley's algorithm} \label{app:2_result-of-Seeley}
This appendix presents several key mathematical equations that are crucial for understanding and implementing Seeley's algorithm. These equations provide the foundational principles and calculations necessary for the algorithm's effective application and analysis. By detailing these equations, this section aims to offer a comprehensive resource for users who wish to gain a deeper understanding of the mathematical underpinnings of Seeley's algorithm.

\subsection{The symbols $\mathbbm{b}_n$} \label{app-results-full-symbol}
To follow the Seeley's algorithm, the symbols $\mathbbm{b}_n$ are needed. To obtain them, one needs to compute them step by step using the recurrence relations \eqref{recurrence-bi} and then represent them as the form \eqref{struct-bn}.
In this appendix, we show the non-zero coefficients appearing on the symbol \eqref{struct-bn} without imposing the Riemann normal coordinate as the special coordinate system.
From the definition of the coefficients \eqref{struct-bn}, it can be seen that they are totally symmetric tensors since we can exchange the indices on the co-vectors $\xi_\mu$. Then, they can be expressed as:
\begin{align}
\begin{aligned}
\mathbbm{b}_{(1)}^{\mu_1}&=
    i \mathbb{H}^{\mu_1}
    \,,\\
\mathbbm{b}_{(1)}^{\mu_1 \mu_2 \mu_3}&=
    -2i {g^{\mu_1 \mu_2}}_{,\lambda} g^{\lambda \mu_3}
    \,,
\end{aligned}
\label{coeff-in-b1}
\end{align}
\begin{align}
\begin{aligned}
\mathbbm{b}_{(2)}^{\bullet}&=
    \mathbb{E}
    \,,\\
%%%
\mathbbm{b}_{(2)}^{\mu_1 \mu_2}&=
    i \mathbbm{b}_{(1)}^{\mu_1} \mathbb{H}^{\mu_2}
    - {g^{\mu_1\mu_2}}_{,\rho\sigma} g^{\rho\sigma}
    + i {g^{\mu_1\mu_2}}_{,\alpha} \mathbbm{b}_{(1)}^\alpha
    - 2 g^{\mu_1\alpha} \partial_\alpha \mathbb{H}^{\mu_2}
    \,,\\
%%%
\mathbbm{b}_{(2)}^{\mu_1 \mu_2 \mu_3 \mu_4}&=
    i \mathbbm{b}_{(1)}^{\mu_1\mu_2\mu_3} \mathbb{H}^{\mu_4}
    + 3 i {g^{\mu_1 \mu_2}}_{,\alpha} \mathbbm{b}_{(1)}^{(\mu_3\mu_4\alpha)}  
% \\&
    - 4 i {g^{\mu_1\mu_2}}_{,\alpha} g^{\alpha\mu_3} \mathbbm{b}_{(1)}^{\mu_4}
    + 4 {g^{\mu_1\mu_2}}_{,\alpha\beta} g^{\alpha\mu_3}g^{\beta\mu_4} \mathbbm{1}_N
    \,,\\
%%%
\mathbbm{b}_{(2)}^{\mu_1 \mu_2 \mu_3 \mu_4 \mu_5 \mu_6}&=
	- 6 i {g^{\mu_1 \mu_2}}_{,\alpha} g^{\alpha \mu_3} \mathbbm{b}^{\mu_4 \mu_5 \mu_6}_{(1)}\,,
\end{aligned}
\label{coeff-in-b2}
\end{align}
\begin{align}
\mathbbm{b}_{(3)}^{\mu_1}&=
	2 i g^{\mu_1 \alpha} \partial_\alpha \mathbb{E}
	+ i g^{\alpha \beta} \partial_\alpha \partial_\beta \mathbb{H}^{\mu_1}
	+ \mathbbm{b}_{(1)}^{\mu_1} \mathbb{E}
	+ i \mathbbm{b}_{(2)}^{\bullet} \mathbb{H}^{\mu_1}
	+ \mathbbm{b}_{(1)}^{\alpha} \partial_{\alpha} \mathbb{H}^{\mu_1}
    \,,\\
%%%
\mathbbm{b}_{(3)}^{\mu_1 \mu_2 \mu_3}&=
    -4i g^{\mu_1 \alpha} g^{\mu_2 \beta} \partial_\alpha \partial_\beta \mathbb{H}^{\mu_3}
    -4i g^{\alpha\beta} g^{\gamma \mu_1} {g^{\mu_2 \mu_3}}_{,\alpha\beta\gamma}
    + \mathbbm{b}_{(1)}^{\mu_1 \mu_2 \mu_3} \mathbb{E}
    + i \mathbbm{b}_{(2)}^{\mu_1 \mu_2} \mathbb{H}^{\mu_3}
\nonumber\\&
    - 4 g^{\mu_1 \alpha} \mathbbm{b}_{(1)}^{\mu_2} \partial_\alpha \mathbb{H}^{\mu_3}
    + 3 \mathbbm{b}_{(1)}^{(\mu_1 \mu_2 \alpha)} \partial_\alpha \mathbb{H}^{\mu_3}
    -4i {g^{\mu_1 \mu_2}}_{,\alpha} g^{\alpha \mu_3} \mathbbm{b}_{(2)}^{\bullet}
    + 2i {g^{\mu_1 \mu_2}}_{,\alpha}\mathbbm{b}_{(2)}^{(\alpha \mu_3)}
 \nonumber\\&
    - 4 {g^{\mu_1 \mu_2}}_{,\alpha\beta} g^{\alpha \mu_3} \mathbbm{b}_{(1)}^\beta
% \nonumber\\&
    - 2 {g^{\mu_1 \mu_2}}_{,\alpha\beta} g^{\alpha\beta} \mathbbm{b}_{(1)}^{\mu_3}
    + 3 {g^{\mu_1 \mu_2}}_{,\alpha\beta} \mathbbm{b}_{(1)}^{(\mu_3 \alpha \beta)}   
    \,,\\
%%%
\mathbbm{b}_{(3)}^{\mu_1 \mu_2 \mu_3 \mu_4 \mu_5}&=
    8 i {g^{\mu_1 \mu_2}}_{,\alpha\beta\gamma} g^{\alpha\mu_3} g^{\beta\mu_4} g^{\gamma\mu_5}
    + \mathbbm{b}_{(2)}^{\mu_1 \mu_2 \mu_3 \mu_4} \mathbb{H}^{\mu_5}
    - 6 g^{\mu_1 \alpha} \mathbbm{b}_{(1)}^{\mu_2 \mu_3 \mu_4} \partial_\alpha \mathbb{H}^{\mu_5}
\nonumber\\&
    - 6 i g^{\mu_1 \alpha} {g^{\mu_2 \mu_3}}_{,\alpha} \mathbbm{b}_{(2)}^{\mu_4 \mu_5}
    + 4 i {g^{\mu_1 \mu_2}}_{,\alpha} \mathbbm{b}_{(2)}^{(\mu_3 \mu_4 \mu_5 \alpha)}
    + 12 {g^{\mu_1 \mu_2}}_{,\alpha\beta} g^{\alpha\mu_3} g^{\beta\mu_4} \mathbbm{b}_{(1)}^{\mu_5}
\nonumber\\&
    - 3 {g^{\alpha\beta}} {g^{\mu_1 \mu_2}}_{,\alpha\beta} \mathbbm{b}_{(1)}^{\mu_3 \mu_4 \mu_5}
% \nonumber\\&
    - 18 {g^{\mu_1 \mu_2}}_{,\alpha\beta} g^{\alpha\mu_2} \mathbbm{b}_{(1)}^{(\mu_4 \mu_5 \alpha)}
      \,,\\
%%%
\mathbbm{b}_{(3)}^{\mu_1 \mu_2 \mu_3 \mu_4 \mu_5 \mu_6 \mu_7}&=
    i \mathbbm{b}_{(2)}^{\mu_1 \mu_2 \mu_3 \mu_4 \mu_5 \mu_6} \mathbb{H}^{\mu_7}
    - 8 i {g^{\mu_1 \mu_2}}_{,\alpha} g^{\alpha \mu_3} \mathbbm{b}_{(2)}^{\mu_4 \mu_5 \mu_6 \mu_7}
    + 6i {g^{\mu_1 \mu_2}}_{,\alpha} \mathbbm{b}_{(2)}^{(\mu_3 \mu_4 \mu_5 \mu_6 \mu_7 \alpha)}
    \,,\\
%%%
\mathbbm{b}_{(3)}^{\mu_1 \mu_2 \mu_3 \mu_4 \mu_5 \mu_6 \mu_7 \mu_8 \mu_9}&=
    - 10 i {g^{\mu_1 \mu_2}}_{,\alpha} g^{\alpha \mu_3}
        \mathbbm{b}_{(2)}^{\mu_4 \mu_5 \mu_6 \mu_7 \mu_8 \mu_9}
    \,,
    %
%%%
\label{coeff-in-b3}
\end{align}
\begin{align}
\mathbbm{b}_{(4)}^{\bullet}&=
	g^{\alpha\beta} \partial_\alpha \partial_\beta \mathbb{E}
	+ \mathbbm{b}_{(2)}^{\bullet} \mathbb{E}
	- i \mathbbm{b}_{(1)}^{\alpha} \partial_\alpha \mathbb{E}
\,,    \\
\mathbbm{b}_{(4)}^{\mu_1 \mu_2}&=
    - 4 g^{\mu_1 \alpha} g^{\mu_2 \beta} \partial_\alpha \partial_\beta \mathbb{E}
    - 4 g^{\alpha\beta} g^{\gamma \mu_1} \partial_\alpha \partial_\beta \partial_\gamma \mathbb{H}^{\mu_2}
    - g^{\alpha\beta} g^{\gamma\lambda} {g^{\mu_1 \mu_2}}_{\alpha\beta\gamma\lambda}
\nonumber\\&\qquad
    + 4 i g^{\mu_1 \alpha} \mathbbm{b}_{(1)}^{\mu_2} \partial_\alpha \mathbb{E}
    - 3i \mathbbm{b}_{(1)}^{(\mu_1 \mu_2 \alpha)} \partial_\alpha \mathbb{E}
    + 2 i g^{\alpha\beta} \mathbbm{b}_{(1)}^{\mu_1} \partial_\alpha \partial_\beta \mathbb{H}^{\mu_2}
\nonumber\\&\qquad
    + 4 i g^{\mu_1 \alpha} \mathbbm{b}_{(1)}^\beta \partial_\alpha \partial_\beta \mathbb{H}^{\mu_2}
    - 3i \mathbbm{b}_{(1)}^{(\mu_1 \alpha \beta)} \partial_\alpha \partial_\beta \mathbb{H}^{\mu_2}
    + 2 i {g^{\mu_1 \mu_2}}_{,\alpha\beta\gamma} g^{\alpha\beta} \mathbbm{b}_{(1)}^\gamma
\nonumber\\&\qquad
    - i {g^{\mu_1 \mu_2}}_{,\alpha\beta\gamma} \mathbbm{b}_{(1)}^{\alpha\beta\gamma}
    + \mathbbm{b}_{(2)}^{\mu_1 \mu_2} \mathbb{E}
    - 4 g^{\mu_1 \alpha} \mathbbm{b}_{(2)}^\bullet \partial_\alpha \mathbb{H}^{\mu_2}
    + 2\mathbbm{b}_{(2)}^{(\mu_1 \alpha)} \partial_\alpha \mathbb{H}^{\mu_2}
\nonumber\\&\qquad
    - 2 g^{\alpha\beta} {g^{\mu_1 \mu_2}}_{,\alpha\beta} \mathbbm{b}_{(2)}^\bullet
    + {g^{\mu_1 \mu_2}}_{,\alpha\beta} \mathbbm{b}_{(2)}^{\alpha\beta}
    + i \mathbbm{b}_{(3)}^{\mu_1} \mathbb{H}^{\mu_2}
    + i {g^{\mu_1 \mu_2}}_{,\alpha} \mathbbm{b}_{(3)}^\alpha\,,
\end{align}
\begin{align}
\mathbbm{b}_{(4)}^{\mu_1 \mu_2 \mu_3 \mu_4}&=
    8 g^{\mu_1 \alpha} g^{\mu_2 \beta} g^{\mu_3 \gamma} \partial_\alpha \partial_\beta \partial_\gamma \mathbb{H}^{\mu_4}
    + 12 g^{\alpha\beta} g^{\gamma\mu_1} g^{\lambda \mu_2} {g^{\mu_3 \mu_4}}_{,\alpha\beta\gamma\lambda} 
    + 6i g^{\alpha \mu_1} \mathbbm{b}_{(1)}^{\mu_2 \mu_3 \mu_4} \partial_\alpha \mathbb{E}
\nonumber\\&\qquad
    - 12i g^{\alpha\mu_1} g^{\beta\mu_2} \mathbbm{b}_{(1)}^{\mu_3} \partial_\alpha \partial_\beta \mathbb{H}^{\mu_4}
    + 3i g^{\alpha\beta} \mathbbm{b}_{(1)}^{\mu_1 \mu_2 \mu_3} \partial_\alpha \partial_\beta \mathbb{H}^{\mu_4}
% \\&\qquad
    + 18i g^{\alpha\mu_1} \mathbbm{b}_{(1)}^{(\mu_2 \mu_3 \beta)} \partial_\alpha \partial_\beta \mathbb{H}^{\mu_4}
\nonumber\\&\qquad
    - 12i g^{\mu_1 \alpha} g^{\beta\gamma} {g^{\mu_2\mu_3}}_{,\alpha\beta\gamma} \mathbbm{b}_{(1)}^{\mu_4}
    - 12i g^{\mu_1 \alpha} g^{\mu_2 \beta} {g^{\mu_3\mu_4}}_{,\alpha\beta\gamma} \mathbbm{b}_{(1)}^{\gamma}
% \\&\qquad
    + 12i g^{\alpha\beta} {g^{\mu_1 \mu_2}}_{,\alpha\beta\gamma} \mathbbm{b}_{(1)}^{(\mu_3 \mu_4 \gamma)}
\nonumber\\&\qquad
    + 18i g^{\mu_1 \alpha} {g^{\mu_2 \mu_3}}_{,\alpha\beta\gamma} \mathbbm{b}_{(1)}^{(\mu_4 \beta \gamma)}
    + i \mathbbm{b}_{(3)}^{\mu_1 \mu_2 \mu_3} \mathbb{H}^{\mu_4}
    - 6 g^{\mu_1 \alpha} \mathbbm{b}_{(2)}^{\mu_2 \mu_3} \partial_\alpha \mathbb{H}^{\mu_4}
% \\&\qquad
    + 4 \mathbbm{b}_{(2)}^{(\mu_1 \mu_2 \mu_3 \alpha)} \partial_\alpha \mathbb{H}^{\mu_4}
\nonumber\\&\qquad
    + 12 g^{\alpha \mu_1} g^{\beta \mu_2} {g^{\mu_3 \mu_4}}_{,\alpha\beta} \mathbbm{b}^\bullet
    - 6 g^{\alpha \mu_1} {g^{\mu_2 \mu_3}}_{,\alpha\beta} \mathbbm{b}_{(2)}^{(\beta\mu_4)}
    - 3 g^{\alpha\beta} {g^{\mu_1 \mu_2}}_{,\alpha\beta} \mathbbm{b}_{(2)}^{\mu_3 \mu_4}
\nonumber\\&\qquad
     -6i {g^{\mu_1 \mu_2}}_{,\alpha} g^{\alpha \mu_3} \mathbbm{b}_{(1)}^{\mu_4}
     + \mathbbm{b}_{(2)}^{\mu_1 \mu_2 \mu_3 \mu_4} \mathbb{E}
% \\&\qquad
     + 6 {g^{\mu_1 \mu_2}}_{,\alpha\beta} \mathbbm{b}_{(2)}^{(\mu_3 \mu_4 \alpha \beta)}
% \\&\qquad
     + 3i {g^{\mu_1 \mu_2}}_{,\alpha} \mathbbm{b}_{(3)}^{(\mu_3 \mu_4 \alpha)}\,,
    \\
\mathbbm{b}_{(4)}^{\mu_1 \mu_2 \mu_3 \mu_4 \mu_5 \mu_6}&=
    - 16 g^{\mu_1 \alpha} g^{\mu_2 \beta} g^{\mu_3 \gamma} g^{\mu_4 \lambda} {g^{\mu_5 \mu_6}}_{,\alpha\beta\lambda\gamma}
    - 24i g^{\mu_1 \alpha} g^{\mu_2 \beta} \mathbbm{b}_{(1)}^{\mu_3 \mu_4 \mu_5} \partial_\alpha \partial_\beta \mathbb{H}^{\mu_6}
\nonumber\\&\qquad
    + 32i g^{\mu_1 \alpha} g^{\mu_2 \beta} g^{\mu_3 \gamma} {g^{\mu_4 \mu_5}}_{,\alpha\beta\gamma} \mathbbm{b}_{(1)}^{\mu_6}
    - 72i g^{\mu_1 \alpha} g^{\mu_2 \beta} {g^{\mu_3 \mu_4}}_{,\alpha\beta\gamma} \mathbbm{b}_{(1)}^{(\mu_5 \mu_6 \gamma)}
\nonumber\\&\qquad
    - 24i g^{\alpha\beta} g^{\gamma \mu_1} {g^{\mu_2 \mu_3}}_{,\alpha\beta\gamma} \mathbbm{b}_{(1)}^{\mu_4 \mu_5 \mu_6}
    + \mathbbm{b}_{(2)}^{\mu_1 \mu_2 \mu_3 \mu_4 \mu_5 \mu_6} \mathbb{E}
    - 8 g^{\mu_1 \alpha} \mathbbm{b}_{(2)}^{\mu_2 \mu_3 \mu_4 \mu_5 } \partial_\alpha \mathbb{H}^{\mu_6}
\nonumber\\&\qquad
    + \mathbbm{b}_{(3)}^{\mu_1 \mu_2 \mu_3 \mu_4 \mu_5} \mathbb{H}^{\mu_6}
    + 24 {g^{\mu_1 \mu_2}}_{,\alpha\beta} g^{\alpha\mu_3} g^{\beta \mu_4} \mathbbm{b}_{(2)}^{\mu_5 \mu_6}
    - 4 {g^{\mu_1 \mu_2}}_{,\alpha\beta} g^{\alpha\beta} \mathbbm{b}_{(2)}^{\mu_3 \mu_4 \mu_5 \mu_6}
\nonumber\\&\qquad
    + 6 \mathbbm{b}_{(2)}^{(\mu_1 \mu_2 \mu_3 \mu_4 \mu_5 \alpha)} \partial_\alpha \mathbb{H}^{\mu_6}
    - 32 {g^{\mu_1\mu_2}}_{,\alpha\beta} g^{\alpha \mu_3} \mathbbm{b}_{(2)}^{(\mu_4 \mu_5 \mu_6 \beta)}
    - 8i {g^{\mu_1\mu_2}}_{,\alpha} g^{\alpha \mu_3} \mathbbm{b}_{(3)}^{\mu_4 \mu_5 \mu_6}
\nonumber\\&\qquad
    + 5 i {g^{\mu_1 \mu_2}}_{,\alpha} \mathbbm{b}_{(3)}^{(\mu_3 \mu_4 \mu_5 \mu_6 \alpha)}
    + 15 {g^{\mu_1 \mu_2}}_{,\alpha\beta} \mathbbm{b}_{(2)}^{(\mu_3 \mu_4 \mu_5 \mu_6 \alpha \beta)}
    \,,
\end{align}
\begin{align}
\mathbbm{b}_{(4)}^{\mu_1 \mu_2 \mu_3 \mu_4 \mu_5 \mu_6 \mu_7 \mu_8}&=
    i \mathbbm{b}_{(3)}^{\mu_1 \mu_2 \mu_3 \mu_4 \mu_5 \mu_6 \mu_7} \mathbb{H}^{\mu_8}
    + 80i g^{\mu_1 \alpha} g^{\mu_2 \beta} g^{\mu_3 \gamma} {g^{\mu_4\mu_5}}_{,\alpha\beta\gamma} \mathbbm{b}_{(1)}^{\mu_6 \mu_7 \mu_8}
\nonumber\\&
    - 10i g^{\mu_1 \alpha} \mathbbm{b}_{(2)}^{\mu_2 \mu_3 \mu_4 \mu_5 \mu_6 \mu_7} \partial_\alpha \mathbb{H}^{\mu_8}
    + 40 g^{\mu_1 \alpha} g^{\mu_2 \beta} {g^{\mu_3 \mu_4}}_{,\alpha\beta} \mathbbm{b}_{(2)}^{\mu_5 \mu_6 \mu_7 \mu_8}
\nonumber\\&
    - 5 g^{\alpha\beta} {g^{\mu_1 \mu_2}}_{,\alpha\beta} \mathbbm{b}_{(2)}^{\mu_3 \mu_4 \mu_5 \mu_6 \mu_7 \mu_8}
    - 10i g^{\mu_1 \alpha} {g^{\mu_2\mu_3}}_{,\alpha} \mathbbm{b}_{(3)}^{\mu_4 \mu_5 \mu_6 \mu_7 \mu_8}
\nonumber\\&
    - 60 g^{\mu_1 \alpha} {g^{\mu_2\mu_3}}_{,\alpha\beta} \mathbbm{b}_{(2)}^{(\mu_4 \mu_5 \mu_6 \mu_7 \mu_8 \beta)}
    + 7 {g^{\mu_1 \mu_2}}_{,\alpha} \mathbbm{b}_{(3)}^{(\mu_3 \mu_4 \mu_5 \mu_6 \mu_7 \mu_8 \alpha)}\,,
    \\
\mathbbm{b}_{(4)}^{\mu_1 \mu_2 \mu_3 \mu_4 \mu_5 \mu_6 \mu_7 \mu_8 \mu_9 \mu_{10}}&=
    i \mathbbm{b}_{(3)}^{\mu_1 \mu_2 \mu_3 \mu_4 \mu_5 \mu_6 \mu_7 \mu_8 \mu_9} \mathbb{H}^{\mu_{10}}
    + 60 g^{\mu_1 \alpha} g^{\mu_2 \beta} {g^{\mu_3 \mu_4}}_{,\alpha\beta}
        \mathbbm{b}_{(2)}^{\mu_5 \mu_6 \mu_7 \mu_8 \mu_9 \mu_{10}}
\nonumber\\&\qquad
    - 12 i {g^{\mu_1 \mu_2}}_{,\alpha} g^{\alpha \mu_3}
        \mathbbm{b}_{(3)}^{\mu_4 \mu_5 \mu_6 \mu_7 \mu_8 \mu_9 \mu_{10}}
% \\&
    + 9 i {g^{\mu_1 \mu_2}}_{,\alpha} \mathbbm{b}_{(3)}^{(\mu_3 \mu_4 \mu_5 \mu_6 \mu_7 \mu_8 \mu_9 \mu_{10} \alpha)}
    \,,
\end{align}
\begin{align}
\mathbbm{b}_{(4)}^{\mu_1 \mu_2 \mu_3 \mu_4 \mu_5 \mu_6 \mu_7 \mu_8 \mu_9 \mu_{10} \mu_{11} \mu_{12}}&=
    -14i {g^{\mu_1 \mu_2}}_{,\alpha} g^{\alpha \mu_3}
        \mathbbm{b}_{(3)}^{\mu_4 \mu_5 \mu_6 \mu_7 \mu_8 \mu_9 \mu_{10} \mu_{11} \mu_{12}}\,.
\end{align}

For computing the trace anomaly for the matter sector, for a positive integer $k$, the symbol $\mathbbm{b}_{2k}$ and its coefficients $\mathbbm{b}_{(2k)}^{\mu_1 \mu_2 \ldots \mu_p}$ are relevant to the one-loop effective action in the $k$-dimensional case via \eqref{ck-gen-series}.
For instance, we should compute these symbols $\mathbbm{b}_{(n)}^{\mu_1 \ldots \mu_p}$ up to the fourth order ($n=4$) when considering the trace anomaly with quantum field theory in four-dimensional spacetime. Therefore, for our purpose, it is sufficient to compute the coefficients up to this order as it is displayed above.
 We stress that the most important term in the asymptotic expansion highly depends on what theory/model we consider.
\subsection{The values on the Riemann normal coordinate} \label{app-results-riemann}
In this appendix, we provide the quantities of different tensors in Riemann normal coordinates. We denote symmetrization as usual~\cite{Muller:1997zk}:
We introduce the symmetrization of tensors by
\begin{align}
A_{(\mu\nu)}
&=\frac{1}{2!} \,
    \qty(
        A_{\mu\nu} + A_{\nu\mu}
    )\,, \\
A_{(\mu\nu\rho)}
&=\frac{1}{3!}
    \qty(
        A_{\mu\nu\rho} + A_{\nu\rho\mu} + A_{\rho\mu\nu}
        + A_{\nu\mu\rho} + A_{\mu\rho\nu} + A_{\rho\nu\mu}
    ) \,,\\
& \vdots \nonumber
\end{align}
Note that the indices surrounded by a vertical bar $|\cdots|$ are not symmetrized.
In what follows, we denote a value of quantity $(\cdots)$ at the point by the subscript zero, $\qty[(\cdots)]_0$.

\subsubsection{Metrics and its derivatives}
The metric and its derivatives in Riemann normal coordinates become:
\begin{align}
\qty[ g_{\mu\nu} ]_0 &=
    \delta_{\mu\nu}
\,,    \\
\qty[ g_{\mu\nu,\lambda} ]_0 &=
    0\,,
    \\
\qty[g_{\mu\nu,\rho\sigma}]_0 &=
    \frac{2}{3} R_{\mu(\rho\sigma)\nu}\,,
    \\
\qty[g_{\mu\nu,\rho\sigma\alpha}]_0 &=
    \nabla_{(\alpha} R_{|\mu|\rho\sigma)\nu}\,,
    \\
\qty[g_{\mu\nu,\rho\sigma\alpha\beta}]_0 &=
    \frac{6}{5}
        \nabla_{(\beta} \nabla_\alpha R_{|\mu|\rho\sigma)\nu}
    + \frac{16}{15}
        R_{\mu(\rho\sigma}^{\phantom{\mu(\rho\sigma}\lambda}
        R_{|\lambda|\alpha\beta)\nu}\,.
\end{align}

\subsubsection{Spin connections and its derivatives}
We note that the third derivative of the spin connection will vanish after taking summation in the %$K_4$
$\mathbb{K}_2$ in Eq.\eqref{eq:expansion-of-heat-kernel-matrix}. So its value on the Riemann normal coordinate is not needed on our computations. The important quantities then become
\begin{align}
\qty[ \omega_\mu ]_0 &=
    0 \,,\\
\qty[ \partial_\alpha \omega_\mu ]_0 &=
    \frac{1}{2} \mathbb{F}_{\alpha\mu}\,,
    \\
\qty[ \partial_\alpha \partial_\beta \omega_\mu ]_0 &=
    \frac{2}{3} \nabla_{(\alpha} \mathbb{F}_{\beta)\mu}\,,
\end{align}
where the tensor $\mathbb{F}_{\mu\nu}$ is the curvature tensor defined by
\begin{align}
\mathbb{F}_{\mu\nu}
=   \partial_\mu \omega_\nu
    - \partial_\nu \omega_\mu
    + \omega_\mu \omega_\nu
    - \omega_\nu \omega_\mu\,.
\end{align}

\subsubsection{The vector $\mathbb{S}^\mu$ and its derivatives}
For higher derivative tensors computed in Riemann normal coordinates, we can symmetrize the indices that emerge on the partial derivative since they commute.
In this symmetrization, the additional curvature terms will appear due to the commutator of the covariant derivatives, yielding the following quantities
\begin{align}
\qty[ \mathbb{S}^\mu ]_0 &=
    \mathbb{S}^\mu\,,
    \\
\qty[ \partial_\alpha \mathbb{S}^\mu ]_0 &=
    \nabla_\alpha \mathbb{S}^\mu\,,
    \\
\qty[ \partial_{\alpha} \partial_{\beta} \mathbb{S}^\mu ]_0 &=
    \nabla_{(\alpha} \nabla_{\beta)} \mathbb{S}^\mu
        + \frac{1}{3} {R^\mu}_{(\beta|\lambda|\alpha)} \mathbb{S}^\lambda\,,
    \\
\qty[ \partial_{\alpha} \partial_{\beta} \partial_{\gamma} \mathbb{S}^\mu ]_0 &=
    \nabla_{(\alpha} \nabla_{\beta} \nabla_{\gamma)} \mathbb{S}^\mu
        + {R^\mu}_{(\alpha|\lambda|\beta} \nabla_{\gamma)} \mathbb{S}^\lambda
        - \frac{1}{2} \nabla_{(\alpha} {R^\mu}_{\beta\gamma)\lambda} \mathbb{S}^\lambda\,.
\end{align}

\subsubsection{Coefficients $\mathbb{E}$, $\mathbb{H}^\mu$ and its derivatives}
The expressions of $\mathbb{E}$, $\mathbb{H}^\mu$ and their derivatives are easily derived by substituting the result of the previous subsection into the definitions of the matrices $\mathbb{H}^\mu$ and $\mathbb{E}$, yielding
\begin{align}
\qty[ \mathbb{E} ]_0 &=
    \mathbb{X}\,,
    \\
\qty[ \partial_\mu \mathbb{E} ]_0 &=
    \nabla_{\mu} \mathbb{X}
    + \mathbb{S}^\lambda \mathbb{F}_{\mu\lambda}
    + \frac{1}{3} \nabla^\lambda \mathbb{F}_{\mu\lambda}
   \,, \\
\qty[ \partial_\mu \partial_\nu \mathbb{E} ]_0 &=
    \nabla_{(\mu} \nabla_{\nu)} \mathbb{X}
    + 2 \nabla_{(\mu} \mathbb{S}^\lambda \mathbb{F}_{\nu)\lambda}
    + \frac{4}{3} \mathbb{S}^\lambda \nabla_{(\mu} \mathbb{F}_{\nu)\lambda}
    - \frac{1}{2} \mathbb{F}_{\lambda(\mu} \mathbb{F}_{\nu)}^{\phantom{\nu)}\lambda}
    - \frac{2}{3} R^{\lambda}_{\phantom{\lambda}(\mu} \mathbb{F}_{\nu)\lambda}
    + g^{\rho\sigma} \qty[\partial_\mu \partial_\nu \partial_\rho \omega_\sigma]_0\,.
\end{align}
\begin{align}
\qty[ \mathbb{H}^\mu ]_0 &=
    2 \mathbb{S}^\mu
 \,,   \\
\qty[ \partial_\alpha \mathbb{H}^\mu ]_0 &=
    2 \nabla_{\alpha} \mathbb{S}^\mu
    - \frac{2}{3} R^\mu_{\phantom{\mu}\alpha}
    + \mathbb{F}_{\alpha}^{\phantom{\alpha}\mu}\,,
    \\
\qty[ \partial_\alpha \partial_\beta \mathbb{H}^\mu ]_0 &=
    2 \qty[
        \nabla_{(\alpha} \nabla_{\beta)} \mathbb{S}^\mu
        + \frac{1}{3} {R^\mu}_{(\beta|\lambda|\alpha)} \mathbb{S}^\lambda
    ]
    + \frac{4}{3} \nabla_{(\alpha} \mathbb{F}_{\beta)\mu}
    - \qty[ \partial_\alpha \partial_\beta \qty( g^{\rho\sigma} \Gamma^{\mu}_{\phantom{\mu}\sigma\rho})]_0\,,
    \\
\qty[ \partial_\alpha \partial_\beta \partial_\gamma \mathbb{H}^\mu ]_0 &=
    2
    \qty[
        \nabla_{(\alpha} \nabla_{\beta} \nabla_{\gamma)} \mathbb{S}^\mu
        + {R^\mu}_{(\alpha|\lambda|\beta} \nabla_{\gamma)} \mathbb{S}^\lambda
        - \frac{1}{2} \nabla_{(\alpha} {R^\mu}_{\beta\gamma)\lambda} \mathbb{S}^\lambda
    ]
\nonumber\\&\qquad
    + \frac{2}{3} \mathbb{F}_{\lambda(\alpha} R^{\lambda\phantom{\beta\gamma)}\mu}_{\phantom{\lambda}\beta\gamma)}
    - \qty[ \partial_\alpha \partial_\beta \partial_\gamma \qty( g^{\rho\sigma} \Gamma^{\mu}_{\phantom{\mu}\sigma\rho})]_0
    + 2 g^{\mu\lambda} \qty[ \partial_\alpha \partial_\beta \partial_\gamma \omega_\lambda ]_0\,.
\end{align}
In the above equations, we have omitted the value of the object $\qty[ \partial_\alpha \partial_\beta \partial_\gamma \omega_\lambda ]_0$ since this term will vanish due to the symmetry of indices.
We additionally introduce the following tensor $\mathbb{K}^{\mu\nu}$ to simplify the coefficients in the symbols:
\begin{align}
\qty[\mathbb{K}^{\mu\nu}]_0
=
    4\qty(\nabla^\mu \mathbb{S}^\nu + \mathbb{S}^\mu \mathbb{S}^\nu)
    + 2 \mathbb{F}^{\mu\nu}
    - \frac{2}{3} R^{\mu\nu} \mathbbm{1}_N\,.
\end{align}

\subsection{Non-zero symbols on Riemann normal coordinate} \label{app-results-symbols-rnc}
By using the Riemann normal coordinate, we can highly reduce the complexity of Seeley's algorithm.
Here, we show the first two nontrivial terms for the heat kernel expansion \eqref{eq:expansion-of-heat-kernel-matrix} in the Riemann normal coordinate, namely:
\begin{align}
\mathbb{K}_1
&=
    \mathbbm{b}_{(2)}^{\bullet}
    + \frac{1}{2!} X_{\alpha_1 \alpha_2}
        \mathbbm{b}_{(2)}^{\alpha_1 \alpha_2}
  \,,  \\
\mathbb{K}_2
&=
    \frac{1}{2!} \mathbbm{b}_{(4)}^{\bullet}
    + \frac{1}{3!} X_{\alpha_1 \alpha_2}
        \mathbbm{b}_{(4)}^{\alpha_1 \alpha_2}
    + \frac{1}{4!} X_{\alpha_1 \alpha_2 \alpha_3 \alpha_4}
        \mathbbm{b}_{(4)}^{\alpha_1 \alpha_2 \alpha_3 \alpha_4}\,.
\end{align}
In particular, $\mathbb{K}_2$ term is made out of the following terms in the symbol $\mathbbm{b}_{4}$:
\begin{align}
%%%
%%%
&
\begin{aligned}
\left.\mathbbm{b}_{(4)}^{\bullet}\right|_{g_{\mu\nu,\lambda}=0}
&=
    g^{\rho\sigma} \partial_{\rho\sigma} \mathbb{E} + \mathbb{H}^\mu \partial_\mu \mathbb{E} + \mathbb{E}^2
\end{aligned}\,,\\
%%%
%%%
&
\begin{aligned}
\left.\mathbbm{b}_{(4)}^{\mu\nu}\right|_{g_{\mu\nu,\lambda}=0}
&=
    - \left\{
        4 g^{\mu\alpha} g^{\nu\beta} \partial_{\alpha\beta} \mathbb{E}
        + 4 g^{\alpha\beta} g^{\mu\lambda} \partial_{\alpha\beta\lambda} \mathbb{H}^\nu
        + g^{\alpha\beta} g^{\rho\sigma} g^{\mu\nu}_{\phantom{\mu\nu},\alpha\beta\rho\sigma}
        + 4 \mathbb{H}^\mu g^{\nu\lambda} \partial_\lambda \mathbb{E}
\right.\\&\qquad\qquad\left.
        + 2 \mathbb{H}^\lambda g^{\alpha\beta} g^{\mu\nu}_{\phantom{\mu\nu},\alpha\beta\lambda}
        + 2 \mathbb{H}^\mu g^{\alpha\beta} \partial_{\alpha\beta} \mathbb{H}^\nu
        + \mathbb{K}^{\mu\nu} \mathbb{E}
        + 4 g^{\mu\alpha} \mathbb{H}^\beta \partial_{\alpha\beta} \mathbb{H}^\nu
\right.\\&\qquad\qquad\left.
        + \mathbb{E} \qty(
            4 g^{\mu\lambda} \partial_\lambda \mathbb{H}^\nu
            + 2 g^{\alpha\beta} g^{\mu\nu}_{\phantom{\mu\nu},\alpha\beta}
            + \mathbb{H}^\mu \mathbb{H}^\nu
        )
        + 2 \mathbb{K}^{(\mu\lambda)} \partial_\lambda \mathbb{H}^\nu
        + 2 \mathbb{K}^{\alpha\beta} g^{\mu\nu}_{\phantom{\mu\nu},\alpha\beta}
\right.\\&\qquad\qquad\left.
        + 2 g^{\mu\lambda} \qty( \partial_\lambda \mathbb{E} ) \mathbb{H}^\nu
        + g^{\alpha\beta} \qty( \partial_{\alpha\beta} \mathbb{H}^\mu ) \mathbb{H}^\nu
        + \mathbb{H}^\mu \mathbb{E} \mathbb{H}^\nu
        + \mathbb{H}^\lambda \qty( \partial_\lambda \mathbb{H}^\mu ) \mathbb{H}^\nu
    \right\}\,,
\end{aligned}\\
%%%
%%%
&
\begin{aligned}
\left.\mathbbm{b}_{(4)}^{\mu\nu\rho\sigma}\right|_{g_{\mu\nu,\lambda}=0}
&=
    8 g^{\mu\alpha} g^{\nu\beta} g^{\rho\lambda} \partial_{\alpha\beta\lambda} \mathbb{H}^\sigma
    + 12 g^{\mu\mu_1} g^{\nu\nu_1} g^{\alpha\beta} g^{\rho\sigma}_{\phantom{\rho\sigma},\alpha \beta \mu_1 \nu_1}
    + 12 \mathbb{H}^\rho g^{\mu\alpha} g^{\nu\beta} \partial_{\alpha\beta} \mathbb{H}^\sigma
\\&\qquad\qquad
    + 12 \mathbb{H}^\rho g^{\alpha\beta} g^{\sigma\lambda} g^{\mu\nu}_{\phantom{\mu\nu},\alpha\beta\lambda}
    + 12 \mathbb{H}^\lambda g^{\rho\alpha} g^{\sigma\beta} g^{\mu\nu}_{\phantom{\mu\nu},\alpha\beta\lambda}
    + 4 g^{\mu\alpha} g^{\nu\beta} g^{\rho\sigma}_{\phantom{\rho\sigma},\alpha\beta} \mathbb{E}
\\&\qquad\qquad
    + 6 \mathbb{K}^{\mu\nu} g^{\rho\lambda} \partial_\lambda \mathbb{H}^\sigma
    + 8 g^{\mu\lambda} g^{\alpha\beta} g^{\rho\sigma}_{\phantom{\rho\sigma},\lambda\beta} \partial_\alpha \mathbb{H}^\nu
    + 8 g^{\rho\alpha} g^{\sigma\beta} g^{\mu\lambda}_{\phantom{\mu\lambda},\alpha\beta} \partial_\lambda \mathbb{H}^\nu
\\&\qquad\qquad
    + 12 g^{\mu\alpha} g^{\nu\beta} g^{\rho\sigma}_{\phantom{\rho\sigma},\alpha\beta} \mathbb{E}
    + 12 g^{\mu\alpha} \mathbb{K}^{(\nu\beta)} g^{\rho\sigma}_{\phantom{\rho\sigma},\alpha\beta}
    + 3 \mathbb{K}^{\mu\nu} g^{\alpha\beta} g^{\rho\sigma}_{\phantom{\rho\sigma},\alpha\beta}
\\&\qquad\qquad
    + 4 g^{\mu\alpha} g^{\nu\beta} \qty( \partial_{\alpha\beta} \mathbb{H}^\rho ) \mathbb{H}^\sigma
    + \mathbb{K}^{\mu\nu} \mathbb{H}^\rho \mathbb{H}^\sigma
    + 2 \mathbb{H}^\mu \mathbb{H}^\nu g^{\alpha\beta} g^{\rho\sigma}_{\phantom{\rho\sigma},\alpha\beta}
\\&\qquad\qquad
    + 4 \qty(
        g^{\alpha\lambda} g^{\beta\tau} g^{\mu\nu}_{\phantom{\mu\nu},\lambda\tau}
        + g^{\mu\lambda} g^{\nu\tau} g^{\alpha\beta}_{\phantom{\alpha\beta},\lambda\tau}
        + 4 g^{\mu\lambda} g^{\alpha\tau} g^{\nu\beta}_{\phantom{\nu\beta},\lambda\tau}
    ) g^{\rho\sigma}_{\phantom{\rho\sigma},\alpha\beta}
\\&\qquad\qquad
    + 4 \mathbb{H}^\mu g^{\nu\lambda} \qty( \partial_\lambda \mathbb{H}^\rho ) \mathbb{H}^\sigma
    + 4 g^{\alpha\beta} g^{\mu\lambda} g^{\rho\sigma}_{\phantom{\rho\sigma},\alpha\beta\lambda} \mathbb{H}^\nu
    + 4 \mathbb{H}^\alpha \mathbb{H}^\nu g^{\mu\beta} g^{\rho\sigma}_{\phantom{\rho\sigma},\alpha\beta}\,.
\end{aligned}
%%%
%%%
\end{align}
In this paper, we have considered only the first two non-trivial coefficients $c_1(x)$ and $c_2(x)$ in the asymptotic expansion.
If we need to find higher coefficients $c_n(x)$ ($n>2$), the algorithm can be continued in the same way.
However, the higher-order coefficients we consider, the more complicated the corresponding computations become.

\section{The integration of the product of covectors} \label{app:1_tensor-X}
In Sec.~\ref{sssec:asymp-expansion}, the tensor $X_{\alpha_1 \alpha_2 \ldots \alpha_p}$ was introduced as the integration of the product of covector $\xi_\mu$ with the Gaussian kernel $e^{-g^{\mu\nu}\xi_\mu \xi_\nu}$.
From the definition, $X_{\alpha_1 \alpha_2 \ldots \alpha_p}$ vanishes when $p$ is odd.
Therefore, the non-trivial contributions of $X_{\alpha_1 \alpha_2 \ldots \alpha_p}$ appears when $p$ is even.
For a 4-dimensional spacetime, we only need the expressions for $p=0,2,4$. In this cases, the tensor $X_{\alpha_1 \alpha_2 \ldots \alpha_p}$ becomes
\begin{align}
X_{\bullet}
&=
    1\,, \\
X_{\alpha_1\alpha_2}
&=
    \frac{1}{2} g_{\alpha_1\alpha_2}\,, \\
X_{\alpha_1 \alpha_2 \alpha_3 \alpha_4}
&= 
    \frac{1}{4} \qty(
        g_{\alpha_1\alpha_2} g_{\alpha_3\alpha_4}
        + g_{\alpha_1\alpha_3} g_{\alpha_2\alpha_4}
        + g_{\alpha_1\alpha_4} g_{\alpha_2\alpha_3}
    )\,,
\end{align}
where $\bullet$ was introduced to denote that the quantity does not have indices (i.e., it is a scalar).
In Ref.~\cite{Obukhov:1983mm}, the generic form of tensor $X_{\alpha_1 \alpha_2 \ldots \alpha_p}$ is introduced and reads as follows
\begin{align}
X_{\alpha_1 \alpha_2 \ldots \alpha_p}
=   \frac{1}{2^{p}}
    \qty[
        \frac{\partial}{\partial z^{\alpha_1}}
        \frac{\partial}{\partial z^{\alpha_2}}
        \cdots
        \frac{\partial}{\partial z^{\alpha_p}}
            \exp\qty(g_{\mu\nu} z^\mu z^\nu)
    ]_{z \to 0}\,.
\end{align}
This equation can be easily obtained from the symmetry of indices of the metric and their products.

\bibliographystyle{utphys}
\bibliography{references}

\end{document}